\documentclass[fleqn,usenatbib]{mnras}

\usepackage[T1]{fontenc}

\usepackage{newtxtext,newtxmath}

\usepackage{graphicx}	
\usepackage{amsmath}	
\usepackage{amssymb}	

\usepackage[usenames,dvipsnames]{color}

\usepackage[modulo,switch]{lineno}

\newcommand{\JJ}{Paper-I}

\title[1D stellar models with interpolated 3D envelopes]{Mending the structural surface effect of 1D stellar structure models with non-solar metallicities based on interpolated 3D envelopes
}

\author[A. C. S. J{\o}rgensen et al.]{
\and
Andreas Christ S{\o}lvsten J{\o}rgensen$^{1}$\thanks{E-mail: acsj@mpa-garching.mpg.de},
Achim Weiss$^{1}$,
George Angelou$^{1}$ and
\and 
V{\'i}ctor Silva Aguirre$^{2}$
\\
$^{1}$Max-Planck-Institut f\"ur Astrophysik, Karl-Schwarzschild-Str. 1, D-85748 Garching, Germany \\
$^{2}$Stellar Astrophysics Centre (SAC), Department of Physics and Astronomy, Aarhus University, Ny Munkegade 120, DK-8000 Aarhus C, Denmark\\
}
\date{Accepted 30.01.2019. Received 30.01.2019; in original form 08.11.2018}

\pubyear{2018}

\begin{document}
\label{firstpage}
\pagerange{\pageref{firstpage}--\pageref{lastpage}}
\maketitle

\begin{abstract}
1D stellar evolution codes employ rudimentary treatments of turbulent convection. For stars with convective envelopes, this leads to systematic errors in the predicted oscillation frequencies needed for asteroseismology. One way of mending these structural inadequacies is through patching, whereby the outermost layers of 1D models are replaced by the mean stratifications from 3D simulations. In order to viably implement this approach in asteroseismic analysis, interpolation throughout precomputed 3D envelopes is required. We present a method that interpolates throughout precomputed 3D envelopes as a function of {\color{black}effective temperature, surface gravity}, and metallicity. We conduct a series of validation tests that demonstrate that the scheme reliably and accurately reproduces the structures of stellar envelopes and apply our method to the Sun as well as two stars observed by \textit{Kepler}. We parameterize the frequency shift that results from patching and show that the functional forms are evolutionary dependent. In addition  we find that neglecting modal effects, such as non-adiabatic energetics, introduces systematic errors in asteroseimically obtained stellar parameters. Both these results suggest that a cautious approach is necessary when utilizing  empirical surface corrections in lieu of patching models. Our results have important implications, particularly for characterizing exoplanet systems, where accuracy is of utmost concern.
\end{abstract}








\section{Introduction}

{\color{black}Asteroseismology, i.e. the study of stellar oscillations \citep{Leighton1960, Ulrich1970}, has proven invaluable for the precise determination of stellar properties as well as for exoplanet research \cite[e.g.][]{jcd2010, Batalha2014, Lebreton2014}. Through the constraints placed on internal structures, both the CoRoT \citep{Baglin2009} and \textit{Kepler} \citep{Borucki2010} space missions have profoundly contributed to the field of stellar physics. Comprehensive reviews of helioseismology \citep{jcd2002} and asteroseismology \citep{Aerts2010, Chaplin2013} outline many of the successes. In this work we focus in particular on solar-like oscillations in the form of $p$-modes. These modes are  acoustic oscillations that are stochastically excited by turbulent convection whereby pressure acts as the restoring force.}

The accuracy of seismically inferred stellar properties strongly depends on the underlying stellar structure models and the interpretation of asteroseismic data is hence subject to any model inadequacies. 
One such deficiency of current 1D stellar evolution codes is the incorrect depiction of the structure of super-adiabatic convective layers, due to a simplistic parametrization of turbulent convection \cite[e.g.][]{Boehm1958}. For stars with convective envelopes, such as solar-like stars, this structural discrepancy leads to a deficient representation of the surface layers, altering the predicted stellar p-mode frequencies. 

The resulting systematic offset between the model frequencies and asteroseismic observations is referred to as the surface effect \citep{Brown1984, jcd1988}. 
The different methods for dealing with  the surface effect lead to different systematic errors when evaluating stellar parameters \citep{Ball2017,Nsamba2018}, since seismic constraints are used to infer these parameters.  
Systematic errors in these parameter estimates affect the conclusion drawn in stellar physics and related or dependent fields, such as exoplanet research or galactic archeology.

In order to mitigate the surface effect, some authors employ a set of frequency ratios suggested by \cite{Roxburgh2003} that have been shown to be less sensitive to the surface layers \citep{Oti2005,Aguirre2011}. Alternatively, many authors rely on empirical or theoretical surface correction relations. A broad variety of these relations can be found in the literature, of which we will refer to the following in the present work: \cite{Kjeldsen2008} suggest a purely empirical {\color{black}power-law} correction relation based on analysis of the present Sun. \cite{Ball2014} present a theoretically motivated relation, based on a asymptotic analysis by \cite{Gough1990}. \cite{Sonoi2015} suggest yet a different functional form based on ten 1D structures, for which the authors have substituted the outermost layers by mean stratifications of 3D hydrodynamical simulations. 

Correcting the structural contribution to the surface effect by substituting the envelope with mean stratifications from hydrodynamic simulations is known as patching and was originally suggested by \cite{Schlattl1997} and \cite{Rosenthal1999}. This was done based on 1D or at most 2D-models. In the last few years, patching has undergone somewhat of a renaissance with several authors adopting this strategy, but using the now available 3D-models \citep{Piau2014, Sonoi2015, Ball2016, Magic2016b, Joergensen2017, Trampedach2017}. However, 3D hydrodynamical simulations are computationally expensive. Rather than computing 3D envelopes that match the restrictions set by a given interior stellar structure, the interior stellar structure was previously chosen in such a way as to match existing 3D simulations. The use of patching has therefore been limited to the analyses of a few special cases, most prominently the present Sun.
Thus, until recently, patching has not been a viable alternative to empirical surface correction relations.

Recently \cite{Joergensen2017} determined a  scheme to interpolate mean stratifications from 3D simulations. This method has proven to accurately mimic the structure of 3D envelopes and has been successfully applied in the analysis of four solar-type dwarfs from the LEGACY and KAGES samples in the \textit{Kepler} field \citep{Aguirre2015, Davies2016, Lund2017,Aguirre2017}. 
This method  allows for the interpolation of the mean 3D-structure in the effective temperature ($T_\mathrm{eff}$) and the gravitational acceleration ($g$) but does not include an interpolation in the composition of the substituted envelope, which limits the applicability to stars with near-solar metallicity.

In this paper, we extend the work by \cite{Joergensen2017}, hereafter \JJ, by additionally interpolating in $\mathrm{[Fe/H]}$. The sparsity of the 3D simulations in this dimension increases the complexity for the original scheme. We show that the method reliably reproduces  the mean structure of 3D simulations throughout the parameter space spanned by $T_\mathrm{eff}$, $\log g$ as well as $\mathrm{[Fe/H]}$. {\color{black}In this paper, we employ the 3D hydrodynamic simulations of stellar envelopes by \cite{Magic2013a}. Hereafter we will refer to these simulations as the Stagger grid. More details are supplied in Section~\ref{sec:stagger}.}

The multidimensional scheme allows us to construct patched stellar models without being restricted to existing 3D simulations. This allows us to map and investigate the structural contribution to the surface effect throughout the parameter space.  
We can express the dependence of the structural surface effect as a function of $T_\mathrm{eff}$, $\log g$ and $\mathrm{[Fe/H]}$ and as per \cite{Sonoi2015} parametrize the structural contribution to the surface effect (sans modal effects, see below). With the current method it is tractable to calculate a vast grid of patched models and determine a surface effect formulation from a well resolved parameter space. 

As we neglect non-adiabatic energetics and do not treat the contribution of the turbulent pressure consistently, we do not expect that the obtained parametrization reproduces observations exactly; the current parameterization considers only the structural contribution of the surface effect --- ignoring, however, the structural contribution that amounts to an adjustment of the adiabatic index by the turbulent pressure. These neglected contributions are collectively referred to as modal effects \citep[cf.][]{Houdek2017}.

In Section~\ref{sec:stagger} and \ref{sec:PatchingProcedure} we discuss the employed grid of 3D simulations and summarize our patching procedure and interpolation scheme. We test the accuracy of the interpolation scheme and investigate the effect of the obtained interpolation error on the model frequencies based on the solar case. In Section~\ref{sec:Surfcorr}, we introduce and compare different parametrizations of the structural surface effect. Using a maximum likelihood estimation method, we use one such parametrization to estimate stellar parameters in Section~\ref{sec:best-fitting}. The obtained results are compared to those determined based on surface correction relations from the literature. Section~\ref{sec:concl} summarizes our conclusions.


\section{The Stagger grid} \label{sec:stagger}
 
We constructed patched models (PMs) based on the Stagger grid: a grid of 3D hydrodynamic simulations of stellar envelopes published by \citet{Magic2013a}. These 3D simulations all stem from \textit{box-in-a-star} type simulations: {\color{black}that is, these simulations only a cover small representative volume of the outermost convective layers. Since the depth of the simulations only amounts to $0.4-10\,\%$ of the total stellar radius \citep{Magic2013a} any sphericity effects as well as changes in $\log g$ are neglected. The envelope models extend from the nearly adiabatic surface layers to the atmosphere, including the photosphere.}

The Stagger grid was computed with a successor of the \citet{Stein1998} {\color{black}finite-difference large-eddy simulation} code, using the Mihalas-Hummer-D\"appen equation of state \citep[MHD-EOS,][]{Hummer1988}, the composition published by \citet{Asplund2009} (AGSS09), and the MARCS line opacities listed in \citet{Gustafsson2008}.

The Stagger grid contains 206 3D envelopes, of which we excluded 6, {\color{black}since their effective temperature strongly oscillates with time, indicating that the simulations are not yet fully relaxed. In addition, one simulations was excluded due to its subadiabtic stratification: the temperature gradient is lower than the adiabatic gradient throughout the simulation (R.~Collet, private communication).}

The grid contains envelopes with $T_\mathrm{eff}$ between $4000$ and $7000\,$K, $\log g$ between $1.5$ and $5.0$ and $\mathrm{[Fe/H]}$ of {\color{black}$-4.0$, $-3.0$, $-2.0$, $-1.0$, $-0.5$, $0.0$ and $0.5$. The spacing in $T_\mathrm{eff}$ and $\log g$ is $500\,$K and $0.5\,$dex, respectively.} 
For $\mathrm{[Fe/H]}\leq -1$ $\alpha$-enhancement has been taken into account: {\color{black}$[\alpha/\mathrm{Fe}] = 0.4\,$dex \citep[][]{Magic2013b}.} 

In our substitution of the outermost layers, we employ the spatial average taken over layers of constant geometric depth and time, in order to conserve hydrostatic equilibrium. We will refer to these averaged structures as $\langle 3\mathrm{D} \rangle$-envelopes. For a discussion on the influence of the averaging on the evaluated stellar oscillation frequencies, we refer to \cite{Magic2016b}.


\section{The patching and interpolation procedures} \label{sec:PatchingProcedure}

Patching refers to the replacement of the outermost layers of 1D stellar models with a $\langle 3\mathrm{D} \rangle$-envelope. Using the nomenclature from \JJ, we refer to the lowermost point of the patched envelope as the patching point. At radii that are larger than the distance from the patching point to the center all mesh points from the un-patched model (UPM) are discarded and replaced by the mean stratification from 3D simulations. The distance from the patching point to the stellar center is determined by interpolation from the structure of the UPM, based on the patching quantity. 

{\color{black}As argued in \JJ, the temperature ($T$), density ($\rho$), total pressure ($P$) and combinations thereof are all suitable choices for the patching quantity. In theory $\Gamma_1$ is also a potential patching quantity. However, in some of the present cases it varies too slowly as function of the distance from the stellar center at the depth at which the patch is performed. Therefore in practice, it is not a suitable patching quantity, contrary to the solar case (cf. \JJ).}

In this paper, we follow many of the steps outlined in \JJ. We match each UPM with a $\langle 3\mathrm{D} \rangle$-envelope of the same effective temperature, gravity and metallicity. {\color{black} We note that other authors take the opposite approach, choosing the interior model in such a way that it matches the the stratification of a given envelope at the patching point \citep[e.g.][]{Ball2016}. This is due to the cost of computing 3D envelopes -- without an interpolation algorithm they are limited to the 
existing 3D simulations. }
{\color{black}
With the interpolation scheme presented in this paper and \JJ, 
it is tractable to reliably construct $\langle 3\mathrm{D} \rangle$-envelopes. This allows us to compute a stellar model with any interior structure that matches observations and then subsequently correct for the shortcomings of the outermost layers, by patching a suitable envelope.
Following this approach permits us to compute PMs with any combination of $T_\mathrm{eff}$, $\log g$ and $\mathrm{[Fe/H]}$ within a wide region of the parameter space rather than being restricted to those cases, for which 3D simulations happen to exist.}

Due to the omission of turbulent pressure and convective back-warming \citep[cf.][]{Trampedach2013,Trampedach2017}, the envelopes predicted by stellar evolution codes are less extended than their 3D counterparts, i.e. the radius of the PMs is larger\footnote{Of course, e.g. large discontinuities in any quantity that enters the determination of the depth from hydrostatic equilibrium may render this statement incorrect.} than the radius of the corresponding UPMs. We define the radius of the PMs, as the distance from the stellar center to the mesh point, at which $T=T_\mathrm{eff}$. Matching the UPM with an envelope that has the same stellar parameters, we find the difference between the radius of the PMs and the radius of the UPMs to be of the order of $< 0.45\%$ for all 315 models that fulfill our selection criteria in Section~\ref{sec:Surfcorr} --- here, we use $\rho$ as the patching quantity. Consequently, the effective temperature of the patched $\langle 3\mathrm{D} \rangle$-envelope is consistent with the Stefan-Boltzmann law within a few degrees. 
An alternative selection of the parameters of the patched envelopes is discussed by \citet{Joergensen2017}.

The gravitational acceleration of each patched envelope is adjusted {\color{black}as a function of depth}, in order to correct for the fact that $\log g$ is assumed to be constant throughout the 3D simulations. {\color{black}As mentioned in Section~\ref{sec:stagger}, the extent of the envelope is small compared to the stellar radius, and consequently these corrections are small.}

The patched $\langle 3\mathrm{D} \rangle$-envelopes are constructed by interpolation in the Stagger grid. If models with the same metallicity as the UPM exist in the Stagger grid, we interpolate only in the $(T_\mathrm{eff},\log g)$-plane. This interpolation scheme makes use of the fact that the structure of the employed 3D simulations show a high degree of homology: first, $\log_{10}(\rho)$, $\log_{10}(T)$, $\Gamma_1$ and $\log_{10}(P)$ are scaled by the corresponding value at a local minimum in $\partial \log \rho / \partial P$, i.e. the density inversion, near the surface. Second, we interpolate the obtained scaled generic stratifications as well as the associated scaling factors. 
{\color{black}For this purpose, we use the scaled pressure as our coordinate --- that is, all quantities are determined for a common set of scaled pressures, requiring interpolation within the mesh of each simulation.} 
Finally, the scaling is inverted. The depth of each mesh point is subsequently established, based on hydrostatic equilibrium.

\begin{figure}
\centering
\includegraphics[width=\linewidth]{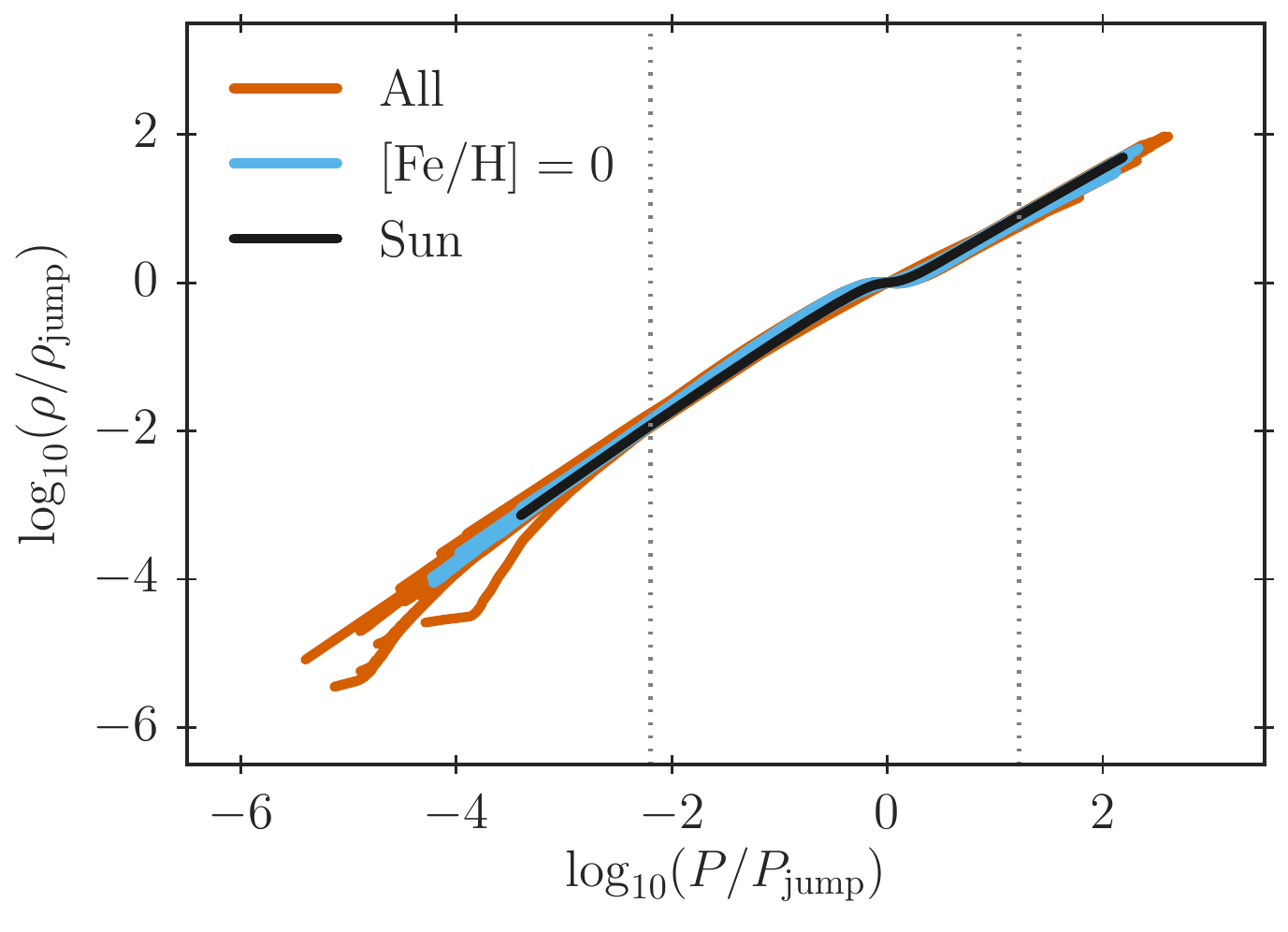}
\caption{The logarithm of the scaled density as a function of the scaled pressure for 199 Stagger-grid envelope models. The scaling factors ($\rho_\mathrm{jump}$ and $P_\mathrm{jump}$) correspond to the value at the density inversion (jump) near the surface. The vertical dotted lines indicate the employed interpolation range.
}
\label{fig:int_range}
\end{figure}

If, on the other hand, no models with the same metallicity exist, we first construct seven envelopes with the desired $T_\mathrm{eff}$ and $\log g$ by interpolation in the $(T_\mathrm{eff},\log g)$-plane. This is performed at each of the seven Stagger grid metallicities: $-4.0$, $-3.0$, $-2.0$, $-1.0$, $-0.5$, $0.0$ and $0.5$. For each of these envelopes, we then rescale $\log_{10}(\rho)$, $\log_{10}(T)$, $\Gamma_1$ and $\log_{10}(P)$ by the corresponding value at the density {\color{black}inversion} near the surface. Based on these scaled logarithmic stratifications, we evaluate $\log_{10}(\rho)$, $\log_{10}(T)$ and $\Gamma_1$ at predetermined $\log_{10}(P)$ for an envelope with the desired metallicity. The interpolation in $\mathrm{[Fe/H]}$ is performed using Akima splines \citep{Akima1970}. Based on the scaling factors of the seven models with the same $T_\mathrm{eff}$ and $\log g$ we determine the scaling factor for the desired parameter values, likewise using Akima spline interpolation. The derivative of the pressure with respect to the density, however, is determined using a third order polynomial spline. The scaling is then inverted. Fig~\ref{fig:int_range} shows the scaled density as function of the scaled pressure for all relaxed models in the grid.

When interpolating in $\mathrm{[Fe/H]}$, the pressure at the deepest point of the interpolated envelopes is a factor of $10^{1.23}$ higher than the pressure at the density jump. This value for the pressure at the deepest point of the interpolated envelope is dictated by the shallowest envelope that enter the interpolation\footnote{We discard the lowermost mesh points of the Stagger-grid envelopes, since we deem these mesh points un-physical: the superadiabatic temperature gradient increases with increasing depth.} (cf.~Fig.~\ref{fig:int_range}). When interpolating in the $(T_\mathrm{eff},\log g)$-plane only, the pressure at the bottom of the envelope may hence be higher. For each PM, we performed the patch at the deepest point of the corresponding interpolated envelopes.  


\subsection{Testing the interpolation scheme}

In order to assess the accuracy of our method we cross validate the interpolation scheme at distinct metallicities in the Stagger grid. We reconstruct 95 Stagger-grid envelopes at $\mathrm{[Fe/H]}$=-3.0, -2.0, -1.0, -0.5 and 0.0. We employ a leave many-out strategy for each atmosphere, withholding all models at the target metallicity. As our splines rely on information on each side of the target metallicity, we are precluded from testing the  metallicities at the edge of the grid, i.e. $\mathrm{[Fe/H]}$=-4.0 and 0.5. These interpolated models are hence determined based on six interpolated models with the same $T_\mathrm{eff}$ and $\log g$ but different $\mathrm{[Fe/H]}$.

The interpolation error at the base of the envelope is of particular interest, as it affects the selection of the patching point. {\color{black}We determined the median, mean and the highest obtained interpolation error at the base for $\Gamma_1(P)$, $T(P)$ and $\rho(P)$, respectively, from our cross-validation exercise and summarize the results in Tab.~\ref{tab:int_error}. Alongside, we list the corresponding values for the errors obtained at any point within the envelopes, as the accuracy varies throughout the structure. Thus, even small displacements of the density inversion will lead to large residuals in the vicinity thereof, due to the large structural changes that take place near the density inversion. The relative interpolation errors are much smaller throughout most of the structures, including the base, than at the density inversion. 

}

\begin{table*}
	\centering
	\caption{Interpolation errors in $\rho$, $T$ and $\Gamma_1$ for each metallicity. {\color{black} The errors listed in the columns denoted by $\rho$, $T$, and $\Gamma_1$ refer to the highest value within the envelopes: the first value of a given entry is the mean of the highest interpolation error within each envelope, the second is the corresponding median, and the third is the highest value among the errors of all envelopes. Consequently, the interpolation errors will be lower than the listed values throughout most of the interpolated structures.} The subscript 'b' refers to the errors of the corresponding quantity at the base of the envelope.}
	\label{tab:int_error}
	\begin{tabular}{lccccccccccccccccccccc} 
		\hline
		$\mathrm{[Fe/H]}$ & $\rho$ [\%] & $\rho_\mathrm{b}$ [\%] & $T$ [\%] & $T_\mathrm{b}$ [\%] & $\Gamma_1$ [\%] & $\Gamma_\mathrm{1,b}$ [\%]  \\
	    \hline
		-3.0 & 6.8, 9.3, 37.0 & 2.3, 3.7, 17.4 & 7.0, 7.7, 16.7  & 0.7, 1.1, 4.2 & 5.8, 6.8, 13.1 & 0.1, 0.3, 2.2 \\
        -2.0 & 10.0, 11.3, 40.1 & 6.6, 7.1, 28.3 & 6.1, 6.3, 12.1 & 1.0, 1.0, 4.7 & 3.9, 4.7, 15.4 & 0.1, 0.2, 0.8 \\
        -1.0 & 11.3, 11.1, 19.7 & 3.5, 3.3, 6.4 & 9.7, 9.1, 14.0 & 1.8, 1.6, 3.0 & 4.0, 4.0, 6.0 & 0.2, 0.3, 0.8 \\
        -0.5 & 8.5, 8.5, 12.1 & 3.6, 3.5, 5.8 & 7.0, 7.5, 13.0 & 1.3, 1.4, 2.0 & 2.7, 3.0, 4.7 & 0.1, 0.2, 0.5 \\
        0.0 & 4.5, 4.6, 7.6 & 0.8, 0.9, 2.2 & 5.7, 5.7, 8.5 & 0.8, 0.8, 1.4 & 2.9, 2.7, 3.7 & 0.1, 0.1, 0.3 \\
        \hline
\end{tabular}
\end{table*}

\begin{figure}
\centering
\includegraphics[width=\linewidth]{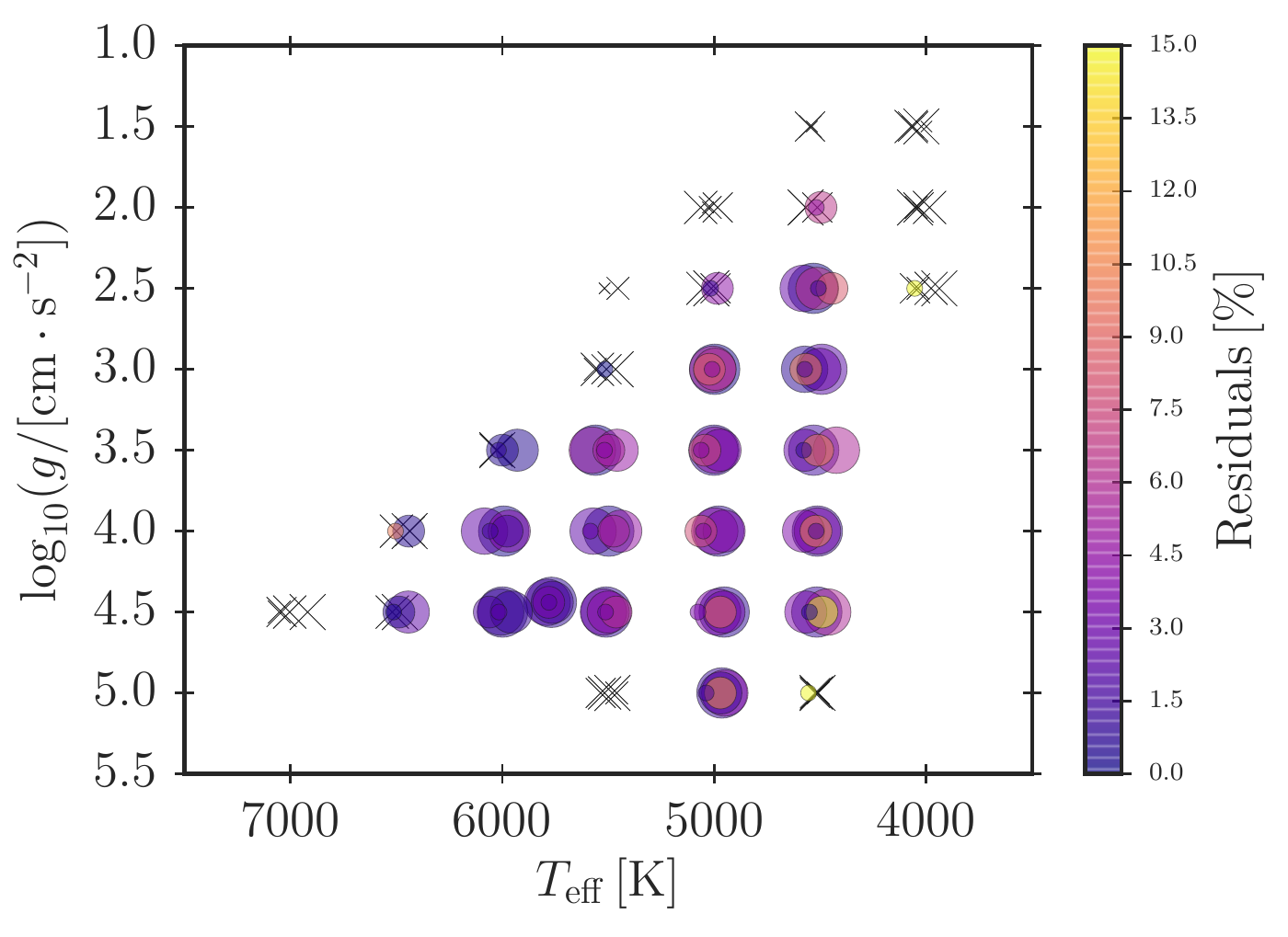}
\caption{Residual of reconstruction of the density at the the bottom of stellar envelopes by interpolation in metallicity. At this mesh point, the pressure is $10^{1.23}$ times higher than at the density inversion near the surface. The larger the marker size is, the larger the metallicity is. The crosses mark the envelopes that could not be reconstructed by interpolation, as they lie on the boundary of the Stagger grid. The color bar shows the obtained errors in percent. Errors higher than $15\,\%$ have all been attributed the same color.
}
\label{fig:error_bottom}
\end{figure}

{\color{black}
The errors at the base of the interpolated envelopes are illustrated in Fig.~\ref{fig:error_bottom}. In none of the 95 cases, do the interpolation error in $\rho(P)$ at the base exceed $28\,\%$. As can be seen from Fig.~\ref{fig:error_bottom}, the interpolation error for the density at the base of the envelope is much lower than the $28\,\%$ throughout most of the Kiel diagram. Indeed, in Fig.~\ref{fig:error_bottom}, the interpolation scheme only performs so badly in the high-$g$ low-$T_\mathrm{eff}$ corner of the grid: the mean and median of the errors shown in Fig.~\ref{fig:error_bottom} are $3.9\,\%$ and $3.0\,\%$, respectively.

Regarding the values stated above, we note that interpolation errors are expected to be smaller than in our cross-validation exercise, i.e. if all models are used. The interpolation errors that enter an analysis of PMs are hence lower than those cited in Tab.~\ref{tab:int_error}.
}

{\color{black}
The highest interpolation errors are obtained for the lowest metallicities. }
For a fixed $\mathrm{[Fe/H]}$, the interpolated values of $T(P)$ and $\rho(P)$ at the base of the envelope are either systematically too high or too low throughout the $(T_\mathrm{eff},\log g)$-plane. Part of this inaccuracy can be attributed to our validation strategy -- we withhold all models at a given metallicity. Nevertheless, our method would benefit greatly from an increase in the Stagger grid resolution.

\begin{figure}
\centering
\includegraphics[width=\linewidth]{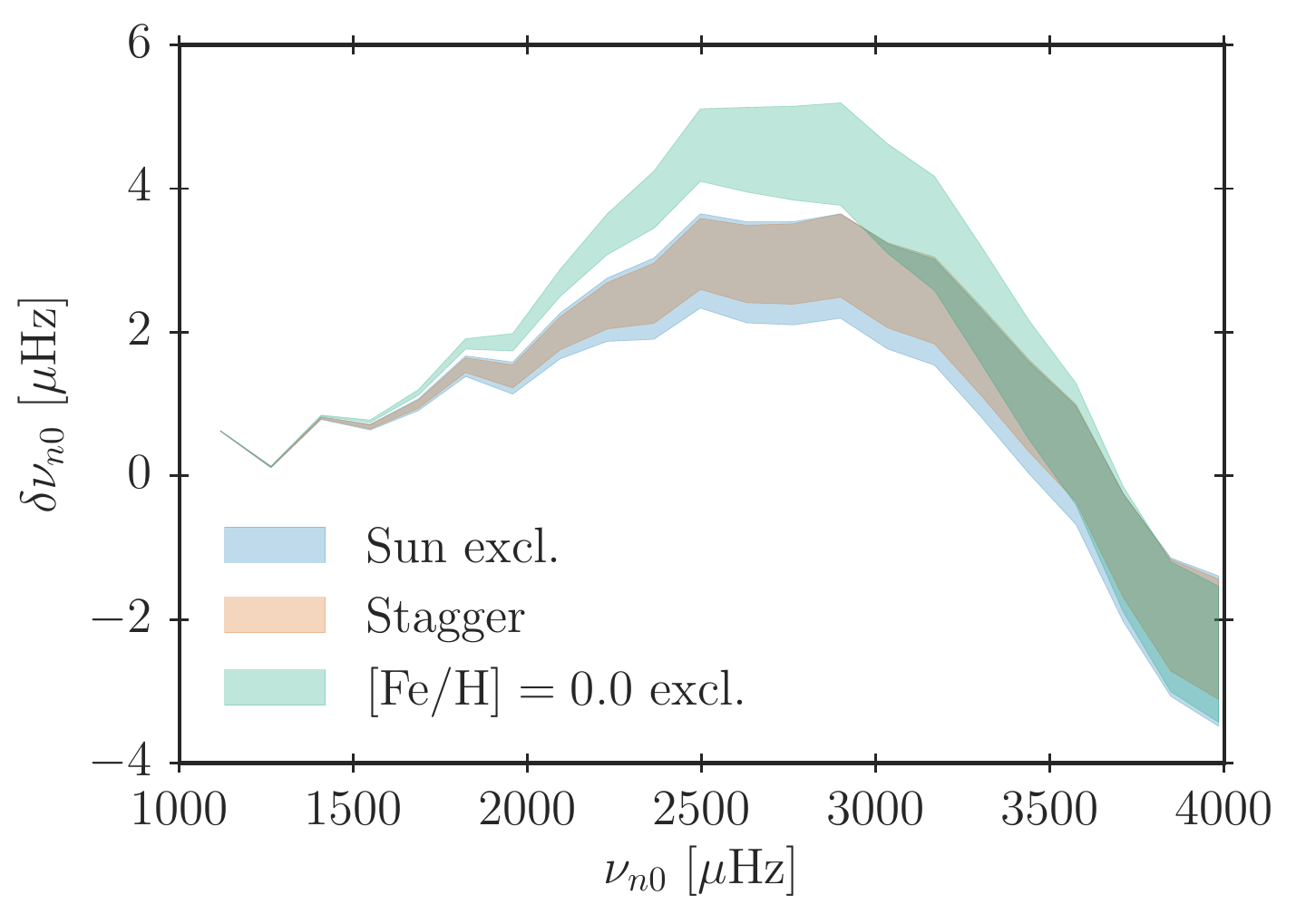}
\caption{Frequency differences between BiSON observations \citep{Broomhall2009,Davies2014} and three PMs. For one model ('Stagger'), we have patched the Stagger-grid envelope with solar global parameters. One model ('Sun excl.') is based on an interpolated envelope that was computed after excluding the solar 3D envelope model from the grid. The last model ('$\mathrm{[Fe/H]}=0$ excl.') has been constructed by patching an interpolated envelope that was computed after excluding all 3D envelopes at solar metallicity from the grid. The width of the shaded area shows the frequency shift that is introduced, when changing the patching patching quantity: we use $P$, $\rho$ or $T$.
}
\label{fig:Sun_Freq}
\end{figure}

To test, whether the interpolation errors affect the predicted model frequencies significantly, we constructed patched solar models. The input physics of the underlying un-patched solar calibration model will be specified in Subsection~\ref{sec:grid1}. For this test, we computed three different envelopes: the original Stagger-grid envelope and two interpolated envelopes. One of the interpolated envelopes was obtained by interpolation in the $(T_\mathrm{eff},\log g)$-plane only after excluding the solar envelope model in the grid. The other interpolated envelope has been computed by interpolation in $T_\mathrm{eff}$, $\log g$ and $\mathrm{[Fe/H]}$, after excluding all models with solar metallicity. We have patched the envelopes to a solar calibration model, using different patching quantities, in order to show the effect of this choice. The resulting model frequency differences, relative to the observed ones, are shown in Fig.~\ref{fig:Sun_Freq}. Note that these differences reach $14\,\mu$Hz at a frequency of $4000\,\mu$Hz for the un-patched solar model. Here, we compare with observations rather than plotting the frequency difference between the UPM and the associated PM to facilitate an easy comparison between Fig.~\ref{fig:Sun_Freq} and the results presented in e.g. \JJ, \cite{Ball2016}, or \cite{Joergensen2018}.

\begin{figure}
\centering
\includegraphics[width=\linewidth]{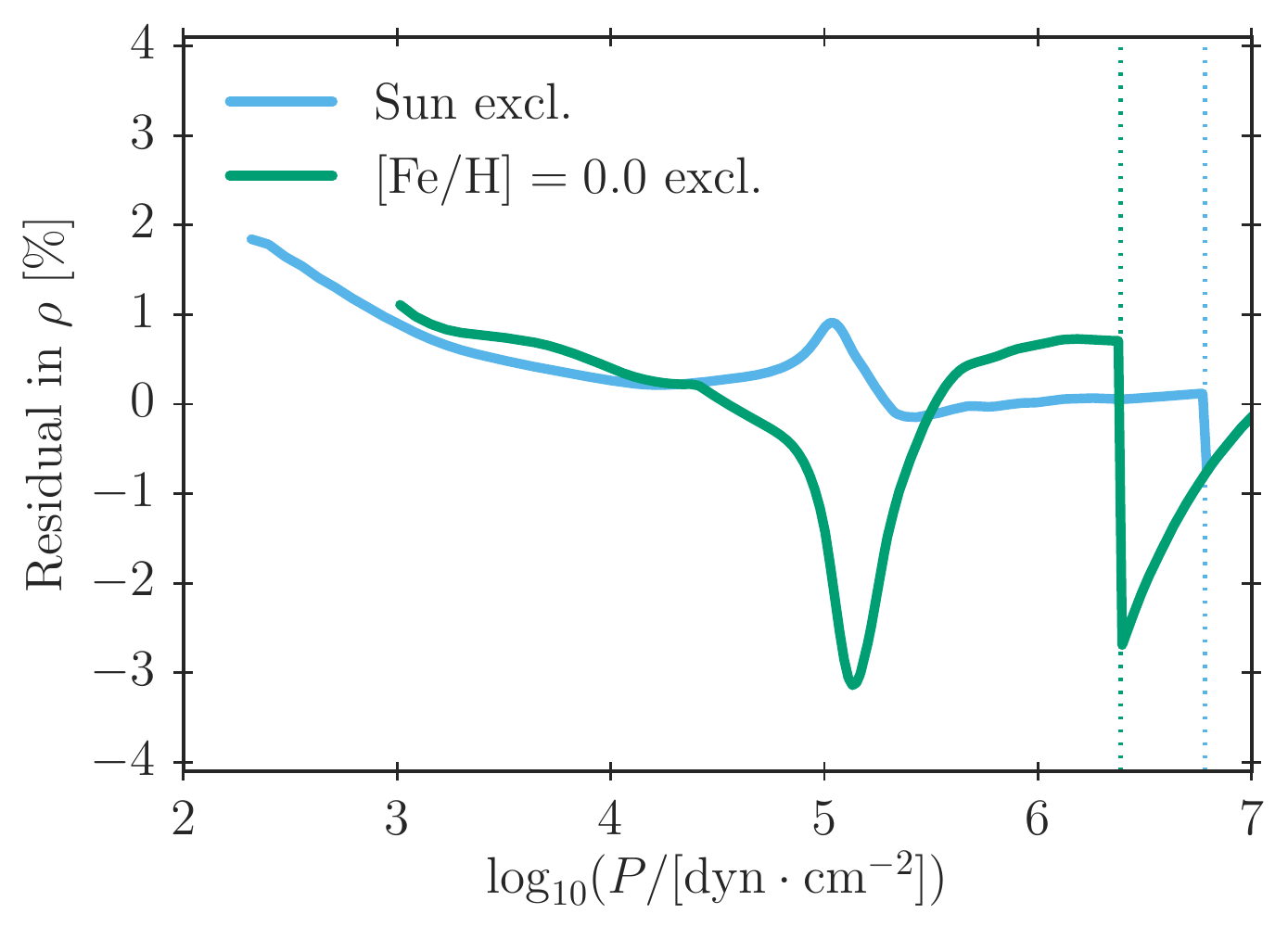}
\caption{Difference between the density of PMs and the solar envelope in the Stagger grid as a function of the total pressure. The vertical dotted lines indicate the location of the patching point. The colors correspond to Fig.~\ref{fig:Sun_Freq}. The density has been used as the patching quantity.
}
\label{fig:Sun_res}
\end{figure}

The structural differences that gives rise to the frequency differences in Fig.~\ref{fig:Sun_Freq} are shown in Fig.~\ref{fig:Sun_res}. Interpolation after the exclusion of all models with solar metallicity leads to a larger misplacement of the density jump near the surface and lower accuracy in general than the interpolation in the $(T_\mathrm{eff},\log g)$-plane. Again, this calls for a refinement of the Stagger grid in $\mathrm{[Fe/H]}$. Furthermore, as can be seen from the figure, the envelope that relies on an interpolation in $\mathrm{[Fe/H]}$ is shallower (with a depth of $1.1\,$Mm) than the envelope obtained from interpolation in only $T_\mathrm{eff}$ and $\log g$ (with a depth of $1.5\,$Mm).
This is because the Stagger-grid envelope with the lowest pressure at its base relative to the pressure at the density inversion dictates the pressure at the base of the interpolated envelopes as described earlier in this section. Due to this, the envelopes employed in this paper are in general slightly shallower than those presented in \JJ: the interpolated envelopes span roughly four orders of magnitude in pressure (cf. Fig.~\ref{fig:int_range}). 

\section{The surface correction} \label{sec:Surfcorr}

We computed a grid of un-patched stellar structure models, using the Garching Stellar Evolution code \citep[\textsc{garstec};][]{Weiss2008}. For each UPM we constructed a PM with the method just described. We then computed stellar oscillation frequencies for all UPMs and PMs in the grid, using the Aarhus adiabatic oscillation package \citep[\textsc{adipls}][]{jcd2008}. 

As argued by \cite{Sonoi2015}, we can use the frequency difference between the UPMs and the associated PMs to derive relations for the {\color{black}surface} correction. We note, however, that the resulting relations cannot be used in the same manner as empirical surface correction relations \citep[e.g.][]{Kjeldsen2008,Ball2014} as we only parametrize the structural contribution.

{\color{black}As can be seen from the solar case illustrated in Fig.~\ref{fig:Sun_Freq}, the oscillation frequencies of the PMs are expected to deviate from the true stellar frequencies. The deviation is only of the order of a few microhertz --- which is why \cite{Sonoi2015} use the adiabatic oscillation frequencies of PMs to approximate the surface effect. However, we note that this statement only holds true, since we use $\Gamma_1$ in the frequency calculations, omitting the corrections introduced to $\Gamma_1$ by the turbulent pressure: turbulent pressure alters the linearized expression that relates the relative gas pressure perturbation and the density perturbation \citep[cf.][]{Rosenthal1999,Houdek2017}. This amounts to reducing $\Gamma_1$ by the ratio of the gas pressure to the total pressure, which leads to an additional shift in the predicted model frequencies \citep[cf.][]{Houdek2017,Joergensen2018}. Altering $\Gamma_1$ accordingly is known as the reduced $\Gamma_1$ approximation, while the use of the unaltered $\Gamma_1$ is known as the gas $\Gamma_1$ approximation. Hence, in accordance with e.g. \cite{Sonoi2015}, \cite{Ball2016}, \cite{Magic2016b}, and \JJ, we employ the gas $\Gamma_1$ approximation.}


\subsection{The stellar model grid} \label{sec:grid1}

We computed evolutionary tracks of un-patched stellar models with masses between $0.65\,\mathrm{M}_\odot$ and $1.50\,\mathrm{M}_\odot$ in steps of $\Delta M = 0.05\,\mathrm{M}_\odot$ from the pre-main sequence (pre-MS\footnote{By which we mean the hydrostatic contraction before the zero age main sequence (ZAMS).{\color{black} Initial pre-MS models are constructed by integration of the stellar
structure equations under the assumption of a constant mass-luminosity relation in the stellar interior, in order to obtain the necessary energy generation term.
This assumption is consistent with a homologous contraction and is dropped as soon as the
evolution proceeds. Pre-MS models typically start at luminosities between $10\, L_\odot$ and $100\, L_\odot$ and central temperatures around $10^5\,\mathrm{K}$.}}) to the red giant branch (RGB) up to a $\log g$ of at least 3.0. We use ten different initial compositions: these include the initial composition obtained from the underlying solar calibration\footnote{More details on a calibration with similar input physics can be found in the paper by \cite{JoergensenWeiss2018}.} as well as initial compositions that correspond to a metallicity, i.e. $\mathrm{[Fe/H]}$, of between -0.3 and 0.5 in steps of 0.1. We restrict ourselves to this metallicity range, as it mostly covers the parameter space that is of interest in connection with \textit{Kepler} targets in the LEGACY and KAGES samples.

For the sake of consistency between the UPMs and the patched Stagger-grid envelopes, we use the solar abundance AGSS09. We employ the corresponding OPAL opacities \citep[][]{Ferguson2005, Iglesias1996} and use the OPAL equation of state \citep[EOS,][]{OPAL2005}, extending the EOS with the EOS by \cite{Hummer1988} at low temperatures. Furthermore, we use the reaction rates suggested by \cite{Adelberger2010} and mixing-length theory for convection \citep{Boehm1958}.  

The solar calibration yields a mixing length ($\alpha_\textsc{mlt}$) of 1.79. {\color{black}We use this value throughout the grid of stellar models. This is a simplifying assumption: it has been established by independent means that $\alpha_\textsc{mlt}$ is not constant with regard to any of the grid parameters \citep{Trampedach2014,Magic2015,Tayar2017}. However, these studies are partly contradictory, i.e. no consensus has been reached on how to correctly vary the mixing length. This being said, it appears that $\alpha_\textsc{mlt}$ varies only moderately in the region of the parameter space that is of interest in our analysis (cf. Figs 4 and 5 in \citealt{Trampedach2014} and Fig.~9 in \citealt{Tayar2017}). Thus, for our purposes, a constant mixing length is a reasonable approximation.
Also, although the assumption of a calibrated mixing length is a simplifying approximation, it is commonly used. Adopting this assumption hence allows a point of comparison between modellers. Furthermore, the main objective of this paper is to demonstrate the usefulness and applicability of the interpolation scheme. This is made all that more tractable by reducing the number of dimensions in our modelling. 

In connection with our choice of $\alpha_\textsc{mlt}$, it is worth mentioning that the structures of PMs show discontinuities at the patching point. This can be seen from the jumps in the residuals in Fig.~\ref{fig:Sun_res}: the density is used as the patching quantity, so while $\rho(r)$ is by construction a smooth function at the patching point, $P(r)$ is not. We also illustrate this in connection with our analysis of \textit{Kepler} stars in Section~\ref{sec:kepler}. These discontinuities result in a sensitivity of the oscillation frequencies to the patching quantity (cf. Fig.~\ref{fig:Sun_Freq}) and reflect inconsistencies between 1D stellar models and 3D envelopes --- the neglect of turbulent pressure or convective back-warming in 1D models are examples of such inconsistencies. For a detailed discussion of this issue, we refer to \JJ. Since $\alpha_\textsc{mlt}$ dictates the adiabat of the stellar model, these discontinuities may partly be mended by calibrating the mixing length anew for each patched model \citep[cf.][]{Ball2016}. However, ensuring smooth stratifications does not guarantee a physically meaningful structure but may rather mask the mentioned inconsistencies. Also, with our method, we aim to correct the outermost layers of stellar models with the desired interior structure rather than to adjust the interior structure to fit the envelope. Instead of artificially ensuring a smooth transition at the patching point, we therefore introduce a set of selection criteria to discard structure models that we do not deem physically reasonable. These criteria are discussed towards the end of this subsection.}

The initial mass fraction of helium ($Y_\mathrm{\odot,i}$) and heavy elements ($Z_\mathrm{\odot,i}$) obtained from this calibration is 0.2653 and 0.0149, respectively. The corresponding model of the present Sun was used in Section~\ref{sec:PatchingProcedure}, when testing the effect of interpolation on the frequency computation (cf. Fig.~\ref{fig:Sun_Freq}).

For bulk compositions different from the Sun, we compute the initial helium mass fraction, using the following linear relation:
\begin{equation}
Y = Y_\mathrm{BBN}+\frac{Y_{\odot,\mathrm{i}}-Y_\mathrm{BBN}}{Z_\mathrm{\odot,i}}Z,
\end{equation}
employing the solar calibration values. $Y_\mathrm{BBN}$ refers to the helium abundance resulting from the Big Bang nucleosynthesis \citep[BBN,][]{Cyburt2016}. We have set $Y_\mathrm{BBN}=0.245$ \citep[cf.][]{Cassisi2003}. In contrast, the mass fraction of helium relative to hydrogen is kept fixed in the Stagger-grid simulations. {\color{black}Specifically,
\begin{equation}
A(\mathrm{He}) = \log_{10}\left(\frac{n_\mathrm{He}}{n_\mathrm{H}} \right) + 12,
\end{equation}
is kept constant. Here, $n_\mathrm{H}$ and $n_\mathrm{He}$ denote the number density of hydrogen and helium, respectively. The relative abundances of the metals correspond to AGSS09 but are scaled to the abundance of Fe (R.~Collet, private communications). As a consequence, the helium abundance in the Stagger-grid envelopes decreases with increasing metallicity, which puts the composition of the 3D simulations somewhat at odds with considerations regarding stellar evolution.}

Thus, while our initial helium mass fraction is reasonable from an evolutionary perspective, it constitutes an inconsistency between the 1D models and 3D envelopes. 
As we aim towards modeling \textit{Kepler} stars in Section~\ref{sec:kepler}, we prefer to include all relevant input physics in order to obtain realistic stellar evolution models rather than artificially mimicking the composition of the 3D simulations. 
The purpose of the following sections is merely to illustrate the applicability and usefulness of our interpolation scheme. While a different set of choices may likewise constitute an interesting test case, we hence restrict ourselves to cases that are relevant for the analysis of \textit{Kepler} stars, ensuring a clean narrative.

From the evolutionary tracks we selected 657 un-patched structure models. All \textsc{garstec} models include microscopic diffusion of H, He, Li, C, N, O, Ne, Mg, Si and Fe. Hence, the surface composition changes during the evolution. 
{\color{black} We note that the incorporation of the microscopic diffusion of metals may lead to an artificially high depletion of metals, especially for low-metallicity models with masses higher than the mass of the Sun. In order to take this into account, we only selected models, for which $[\mathrm{Fe/H}]\geq -0.5$, hereby excluding any model, for which the abundance of heavy elements were vastly depleted.
This issue could have been avoided by ignoring microscopic diffusion, when constructing the grid. However, as stated above, we prefer to include all relevant input physics.
}

{\color{black}As mentioned in Section~\ref{sec:stagger},} the composition of the Stagger-grid envelopes takes $\alpha$-enhancement into
account for $[\mathrm{Fe/H}]\leq -1$. We restrict the grid to metallicities, where no $\alpha$-enhancement is needed, in order for the bulk composition to be trivially related to the metallicity:
\begin{equation}
\mathrm{[Fe/H]}=\log_{10} \left( \frac{Z_\mathrm{S}}{X_\mathrm{S}} \right)-\log_{10} \left( \frac{Z_{\odot,\mathrm{S}}}{X_{\odot,\mathrm{S}}} \right).
\end{equation}
Here, $Z_{\odot,\mathrm{S}}/X_{\odot,\mathrm{S}}=0.01792$, where $Z_\mathrm{S}$ and $X_\mathrm{S}$ denote the mass fraction of metals and hydrogen at the surface respectively.

For each of the 657 UPMs, we constructed PMs, using the density as the patching quantity. As discussed in \JJ, the frequencies are relatively insensitive to the patching depth, if the patch is performed sufficiently deep within the nearly adiabatic region. In order to ensure that the patched envelope is indeed deep enough, we impose the criterion that the patching point must be placed deeper than the minimum in $\Gamma_1$ near the surface. This criterion leads to the omission of 25 models in the {\color{black}high}-$g$, low-$T_\mathrm{eff}$ corner of the Kiel diagram --- {\color{black}that is, main-sequence stars with $\log g = 4.5$ and $T_\mathrm{eff}<5300\,$K.}

We excluded 47 additional models, for which the radius of the UPM exceeds the radius of the associated PM: 3D envelopes are more extended than their 1D counterparts. If the radius of the PM is smaller than the radius of the UPM, this must be associated with a large discontinuity in either $P$ or $\rho$. {\color{black}For all these PMs, $4.0 < \log g <4.3$.} 

As shown in \JJ, the radius, at which the patching point is placed, is sensitive to the choice of the patching quantity. Hence, the patching quantity affects the model frequencies as shown in Fig~\ref{fig:Sun_Freq}. Since we investigate the structural surface effect, relying therefore on the model frequencies, this ambiguity caused by the choice of the patching quantity is undesirable. To take this into account, we constructed two additional PMs for each of the 585 UPMs that have so far successfully passed our selection criteria, using $P$ and $T$ as the patching quantity, respectively. We then excluded 270 models, for which the choice between $P$, $T$ and $\rho$ as the patching quantity affects the model frequencies by more than $2\times 10^{-3} \nu_\mathrm{max}$, unless the deviation fell below $1.0\,\mu$Hz. For the Sun, $2\times 10^{-3} \nu_\mathrm{max}$ would roughly correspond to twice the highest obtained deviation. {\color{black}The models that are hereby discarded are scattered across the Kiel diagram.}

{\color{black}All in all, our selection criteria completely deplete certain regions of the Kiel diagram: all models with $T_\mathrm{eff} \gtrapprox 6000\,$K were discarded as were most cold main sequence stars with $\log g \approx 4.3$. Meanwhile, no PM with $\log g < 3.15$ was affected by any of our selection criteria.}

It is worth to take a closer look at the models that fail to pass the last selection criterion: for some of these models, the patching procedure did not alter the frequencies significantly, which may suggest that the patched envelopes are too shallow in these cases. Our analysis thus calls for deeper 3D simulations --- ideally, new simulations should extend into the adiabatic layers. Furthermore, all 47 models with $R_\mathrm{PM}<R_\mathrm{UPM}$ would have fallen into this category if not previously excluded.

Alternatively, one may require a patching point deeper within the adiabatic region by excluding all models, for which $|\nabla-\nabla_\mathrm{ad}|$ exceeds a certain value. Also, one may exclude all models, for which the choice between $P$, $T$ and $\rho$ as the patching quantity affects the radius of the patching point by more than a certain fraction of the patching depth. In the case of the Sun, this choice affects the radius of the patching point by $21-27\,$km, corresponding to roughly $2-3\,\%$ of the patching depth. However, each additional selection criterion introduces a selection bias: some regions of the Kiel diagram are more harshly depleted than others. As illustrated in Subsection~\ref{sec:SolRec}, such biases will leave their mark on the parameterization of the structural surface effect. We therefore avoid to introduce further selection criteria. At the end, 315 models passed our selection criteria. Due to microscopic diffusion, the surface metallicity of these models range from -0.49 and 0.45.


\subsection{The surface correction relation}

In this paper, we present a parametrization of the structural contribution to the surface effect. We derive surface correction relations based on the frequency difference, 
\begin{equation}
\delta \nu = \nu_\mathrm{PM}-\nu_\mathrm{UPM},
\end{equation}
between the model frequencies of the UPM ($\nu_\mathrm{UPM}$) and the model frequencies of the associated PM ($\nu_\mathrm{PM}$). As we do not consider modal effects, there is no reason to believe that PMs encapsulate the full surface term. We in fact see this in solar case in Fig.~\ref{fig:Sun_Freq}, where  we do not capture all the features of $\delta \nu$. We investigate how well our parameterization recovers the determination of the stellar parameters in Section~\ref{sec:best-fitting} in a comparison with \cite{Ball2014}. 

We use the same general functional forms for the surface correction relations as \cite{Sonoi2015}. This includes a power-law fit to the frequency correction:
\begin{equation}
\frac{\delta \nu}{\nu_\mathrm{max}} =  a \left(\frac{\nu_\mathrm{PM}}{\nu_\mathrm{max}} \right)^b \label{eq:poly}.
\end{equation}
We alternatively try a Lorentzian fit of the form
\begin{equation}
\frac{\delta \nu}{\nu_\mathrm{max}} =  \alpha \left(1-\frac{1}{1-(\nu_\mathrm{PM}/\nu_\mathrm{max})^{\beta}} \right). \label{eq:lorentz}
\end{equation}
In the above relations, $\nu_\mathrm{max}$ denotes the frequency of maximum power. We adopt \citep{Brown1991}
\begin{equation}
\nu_\mathrm{max} = \frac{g}{g_{\odot}}\left(\frac{T_\mathrm{eff}}{T_\mathrm{eff\odot}}\right)^{-1/2}\nu_\mathrm{max\odot}, 
\end{equation}
where $g$ and $T_\mathrm{eff}$ denote gravitational acceleration and the effective temperature of the UPM, respectively. In this paper, we set $\nu_\mathrm{max\odot} = 3090\,\mu$Hz, $T_\mathrm{eff\odot}=5779.57\,$K, and $g_\odot=4.438$ --- that is $R_\odot=6.95508\times10^8\,$cm, $M_\odot=1.9891\times10^{33}\,$g. Finally, $a$, $b$, $\alpha$ and $\beta$ are fitting parameters.

Eq.~(\ref{eq:poly}) only yields a reasonable description of the frequency difference at frequencies that are lower than or close to $\nu_\mathrm{max}$ \citep{Sonoi2015}. While all frequencies are included, when determining the fitting parameters in connection  with Eq.~(\ref{eq:lorentz}), we only include frequencies that are lower than $1.05\nu_\mathrm{max}$, when using Eq.~(\ref{eq:poly}), in accordance with \cite{Sonoi2015}. In either case, we perform the fit, using only radial modes ($\ell = 0$), in accordance with \cite{Kjeldsen2008} and \cite{Sonoi2015}. We use all radial orders {\color{black}in the relevant frequency range}.

Both Eq.~(\ref{eq:poly}) and Eq.~(\ref{eq:lorentz}) have two fitting parameters. These are assumed to be functions of the global stellar parameters. For all four fitting parameters we adopt power law dependences that include four free parameters, e.g.:
\begin{equation}
a = A\left(\frac{T_\mathrm{eff}}{T_\mathrm{eff\odot}}\right)^B\left(\frac{g}{g_{\odot}}\right)^C\left(\frac{Z_\mathrm{S}/X_\mathrm{S}}{Z_\mathrm{\odot,S}/X_\mathrm{\odot,S}} \right)^D. \label{eq:aABCD}
\end{equation}
Again, all values are taken from the respective UPM. Eq.~(\ref{eq:aABCD}) only deviates from the assumed functional dependence adopted by \cite{Sonoi2015} by the dependence on $Z_\mathrm{S}/X_\mathrm{S}$, i.e. on $\mathrm{[Fe/H]}$, since we consider models with non-solar metallicities.

{\color{black}We note that \cite{Kjeldsen2008} have suggested a power-law correction, i.e. a relation on the form of Eq.~(\ref{eq:poly}), using a different definition of the coefficients, $a$ and $b$. We will state this definition in Section~\ref{sec:best-fitting}, where we include a comparison to \cite{Kjeldsen2008}. In this Section, however, all coefficients in Eq.~(\ref{eq:poly}) and Eq.~(\ref{eq:lorentz}) --- that is, $a$, $b$, $\alpha$ and $\beta$ --- are defined equivalently to $a$ in Eq.~(\ref{eq:aABCD}). 
}

\begin{figure}
\centering
\includegraphics[width=\linewidth]{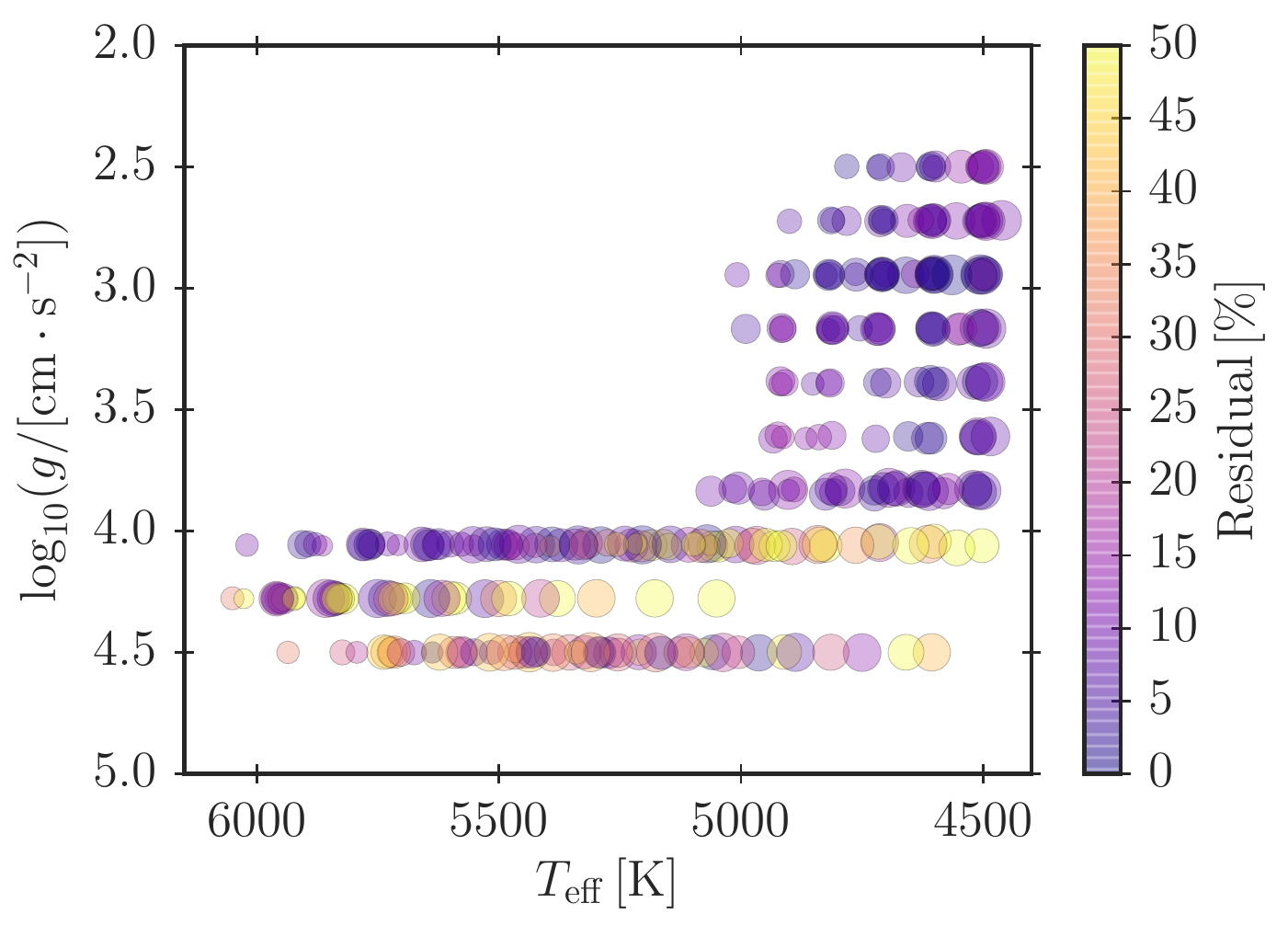}
\caption{Residual of reconstruction of $\delta \nu$ relative to the actual value of $\delta \nu$ for the Lorentzian fit, i.e. Eq.~(\ref{eq:lorentz}), summarized in Table~\ref{tab:fitval}. {\color{black}The residuals are computed at $1.3\nu_\mathrm{max}$  --- for the Lorentzian fit, this lies within the covered frequency range.}
Residuals that exceed $50\,\%$ are indicated by the same color: the highest obtained error is 317$\,\%$. The larger the marker size is, the larger the metallicity is.
}
\label{fig:errorplane}
\end{figure}

It follows from the above that a fit of $\delta \nu$ to either Eq.~(\ref{eq:poly}) or Eq.~(\ref{eq:lorentz}) in a parameter space spanned by $T_\mathrm{eff}$, $\log g$, $\mathrm{[Fe/H]}$ and $\nu/\nu_\mathrm{max}$ includes eight free parameters. In this paper we perform global fits to the entire grid presented in the previous subsection, constraining all eight parameters simultaneously. The advantage of performing such a global fit is that this approach allows for a robust determination of the uncertainties on the predicted frequency corrections based on the variance-covariance matrix ($\mathbfss{C}$) obtained from the fit: for a fixed scaled frequency $\nu_k/\nu_\mathrm{max}$ the associated frequency shift ($\delta \nu_k/\nu_\mathrm{max}$) is a function $f$ of eight free parameters ($\xi_1,...,\xi_8$) given by either Eq.~(\ref{eq:poly}) or Eq.~(\ref{eq:lorentz}) in combination with equations for $a$, $b$, $\alpha$ and $\beta$ of the form of Eq.~(\ref{eq:aABCD}), which provides the Jacobian $\mathbfit{J}$. The variance ($\sigma^2_k$) of $\delta \nu_k/\nu_\mathrm{max}$ is hence
\begin{equation}
\sigma^2_{k} = \mathbfit{J} \mathbfss{C} \mathbfit{J}^\mathrm{T}, \quad 
\mathbfit{J} = \left[\partial f / \partial \xi_1, ... , \partial f/\partial \xi_8 \right]. \label{eq:jakobian}
\end{equation}
The results of our global fit including all 315 PMs and their associated UPMs are summarized in Table~\ref{tab:fitval}. How well the Lorentzian fit reproduces the correct frequency shift throughout the Kiel diagram is furthermore illustrated in Fig.~\ref{fig:errorplane}. 

\begin{table*}
	\centering
	\caption{Fitting parameters based on a global fit to all 315 PMs that fulfilled our selection criteria. {\color{black}Each fitting parameter ($a$, $b$, $\alpha$ and $\beta$) in Eq.~(\ref{eq:poly}) and Eq.~(\ref{eq:lorentz}) is itself composed of four fitting parameters ($A$, $B$, $C$ and $D$) as given in Eq.~(\ref{eq:aABCD})}. The uncertainties denote the square root of the diagonal of the variance-covariance matrix. We have performed the fit, using only radial modes ($\ell = 0$).}
	\label{tab:fitval}
	\begin{tabular}{lccccccccccccccccccccc} 
		\hline
		  & A & B & C & D\\		
		\hline
$a$ &$(-2.259 \pm 0.050)\times 10^{-3}$ & $2.808\pm 0.180$ & $(-7.189\pm 0.089)\times 10^{-1}$ &  $(2.513\pm  0.127)\times 10^{-1}$  \\  
$b$ &$4.517\pm 0.175$ & $-1.290\pm 0.308$ & $(2.020\pm 0.151)\times 10^{-1}$ & $(-1.504\pm  0.214)\times 10^{-1}$\\     
$\alpha$ &$(-3.949\pm 0.035)\times 10^{-3}$ & $2.659\pm 0.071$ & $(-6.888\pm 0.035)\times 10^{-1}$ &  $(2.633\pm 0.052)\times 10^{-1}$   \\
$\beta$ &$5.530\pm  0.153$ & $-1.445\pm 0.217$ & $(2.271\pm 0.106)\times 10^{-1}$ & $(-1.334\pm 0.153)\times 10^{-1}$  \\		\hline
	\end{tabular}
\end{table*}

As can be seen from Fig.~\ref{fig:errorplane}, the fit does not perform equally well throughout the parameter space. {\color{black}Whilst the varying accuracy of an imperfect fit is not surprising, it is worth taking a closer look at this, since it suggests that the surface corrections by \cite{Kjeldsen2008} and \cite{Sonoi2015} are subject to a selection bias: after all, the relation by \cite{Kjeldsen2008} is calibrated based on the Sun only, and the relation by \cite{Sonoi2015} is based on ten PMs, for which $T_\mathrm{eff}$ is predominantly larger than $6000\,$K.
}

If the fit is performed based on a subsample, excluding further models from the grid, we obtain different fitting parameters. Concretely, if we exclude, say, all RGB stars before evaluating the fitting parameters, we obtain a fit that performs better for main sequence (MS) stars and worse for RGB stars. The ramifications of this sensitivity to the subsample selection are illustrated in the next subsection, where we reconstruct the present Sun.


\subsection{Reconstructing the Sun} \label{sec:SolRec}

As a validation for our global fit, we determine how well the parameterization  mimics the  
frequency difference between the UPM and the associated PM for the present Sun. The solar calibration model of the present Sun, based on which we computed the evolution paths that enter the grid, is itself not in the sample of structure models used to evaluate the fitting parameters. This makes it a suitable test case.

\begin{figure}
\centering
\includegraphics[width=\linewidth]{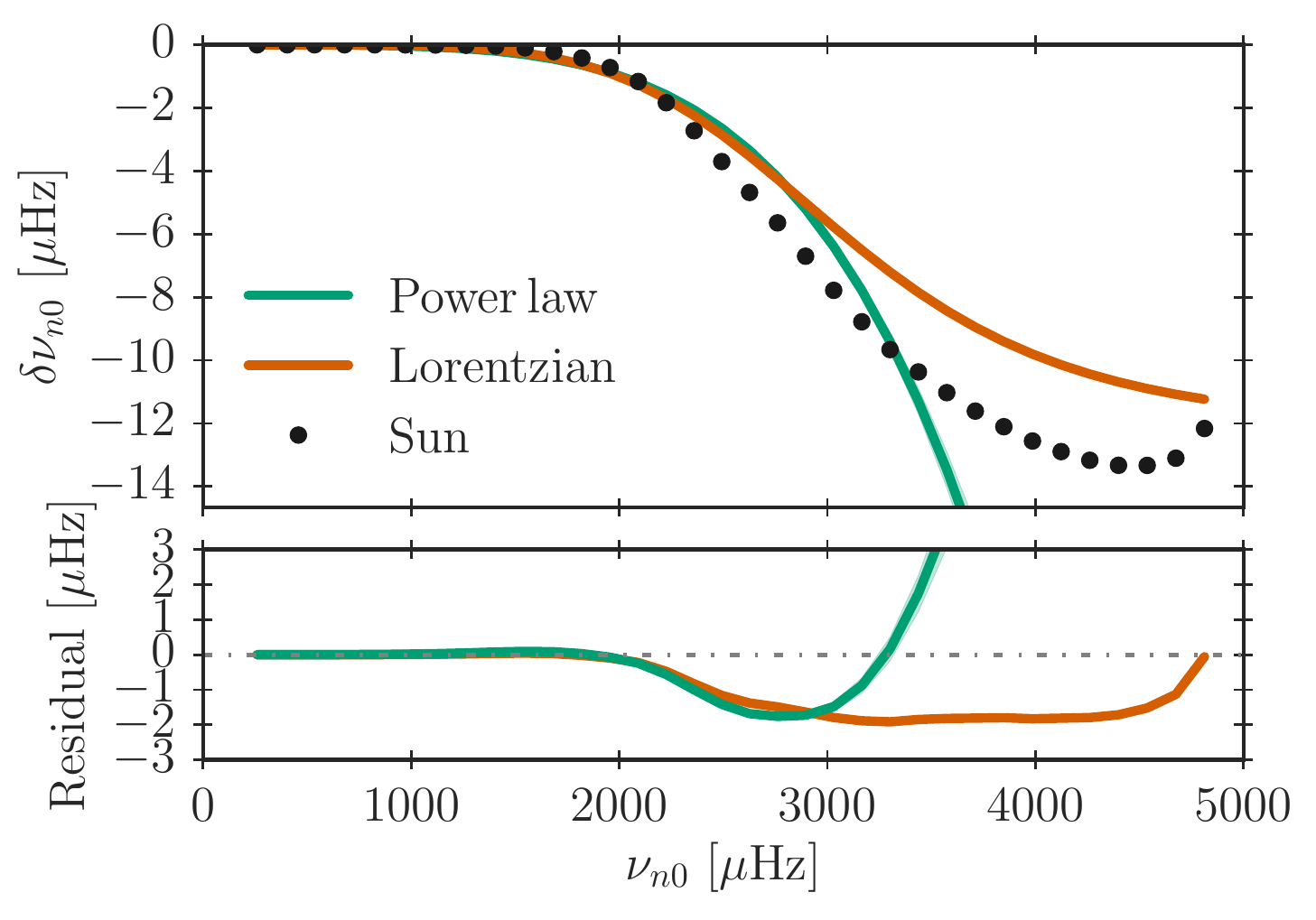}
\caption{Reproduction of the frequency difference between the solar calibration model and the associated PM, using the power-law and Lorentzian fit, i.e. Eqs~(\ref{eq:poly}) and (\ref{eq:lorentz}), respectively. The fits have been evaluated based on all successful PMs in the grid (cf.~Table~\ref{tab:fitval}). The actual frequency difference between the UPM and the PM are labeled "Sun" and marked with circles. The power-law fit is based on low frequencies only.
}
\label{fig:poly_lorentz}
\end{figure}

We have used the fit summarized in Table~\ref{tab:fitval} to predict the frequency difference, employing the solar effective temperature, metallicity and gravitational acceleration. A comparison to the actual frequency difference between the UPM and the PM is shown in Fig.~\ref{fig:poly_lorentz}. Both fits {\color{black}lead to} residuals of $\lessapprox 2\,\mu$Hz within the relevant frequency ranges: the power-law fit is expected to work well up to $\nu_\mathrm{max}$, while the Lorentzian fit performs well up to the acoustic cut-off frequency. {\color{black}Thus, both fits perform rather well, encapsulating most of the structural surface effect --- however, we note that the residuals are still much larger than the typical uncertainties of the observed frequencies.}

\begin{figure}
\centering
\includegraphics[width=\linewidth]{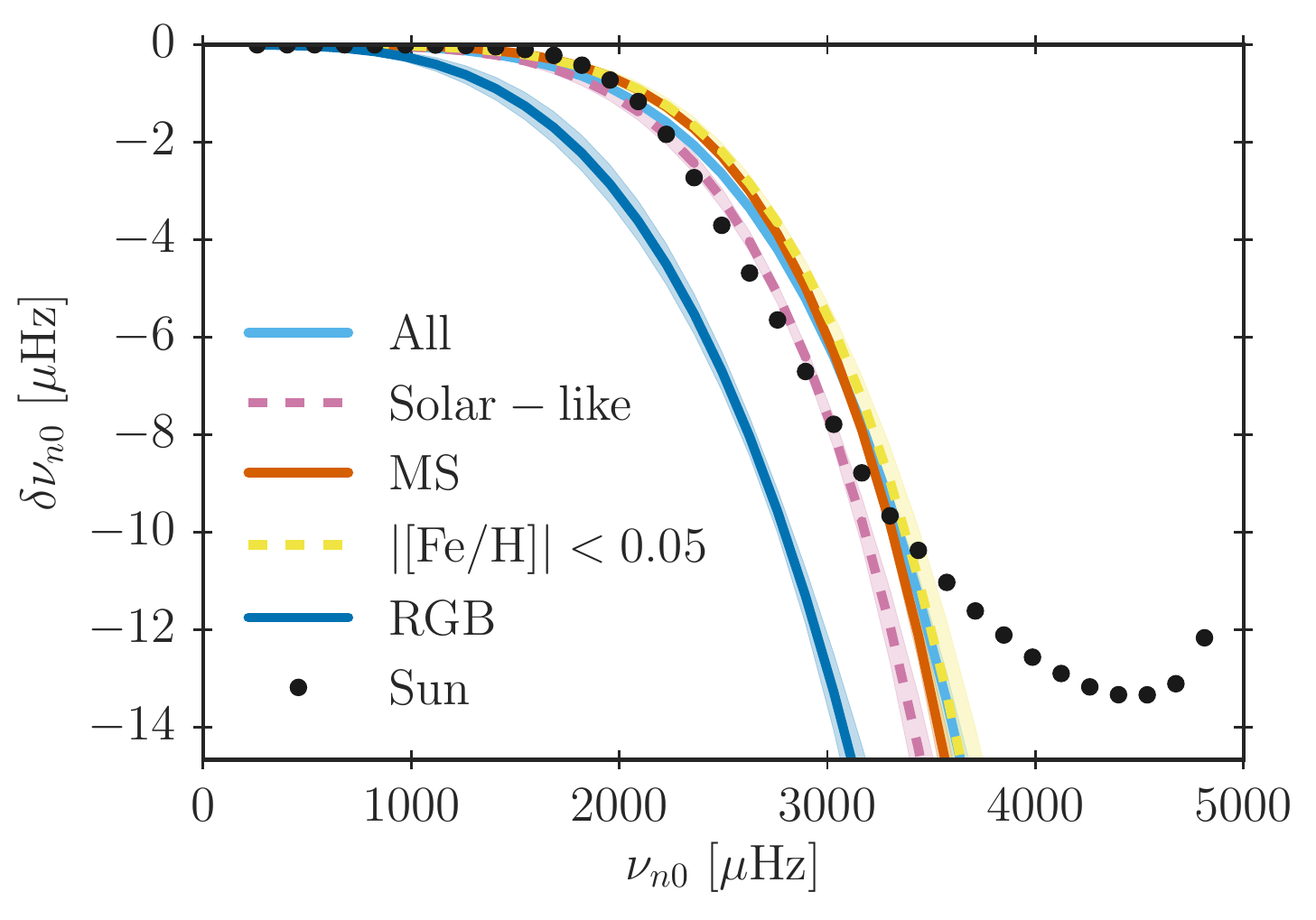}
\caption{Reproduction of the frequency difference between the solar calibration model and the associated PM, using power-law fits to different subsamples of the grid. Details are specified in the text and in the caption of Fig.~\ref{fig:poly_lorentz}.
}
\label{fig:poly}
\end{figure}

\begin{figure}
\centering
\includegraphics[width=\linewidth]{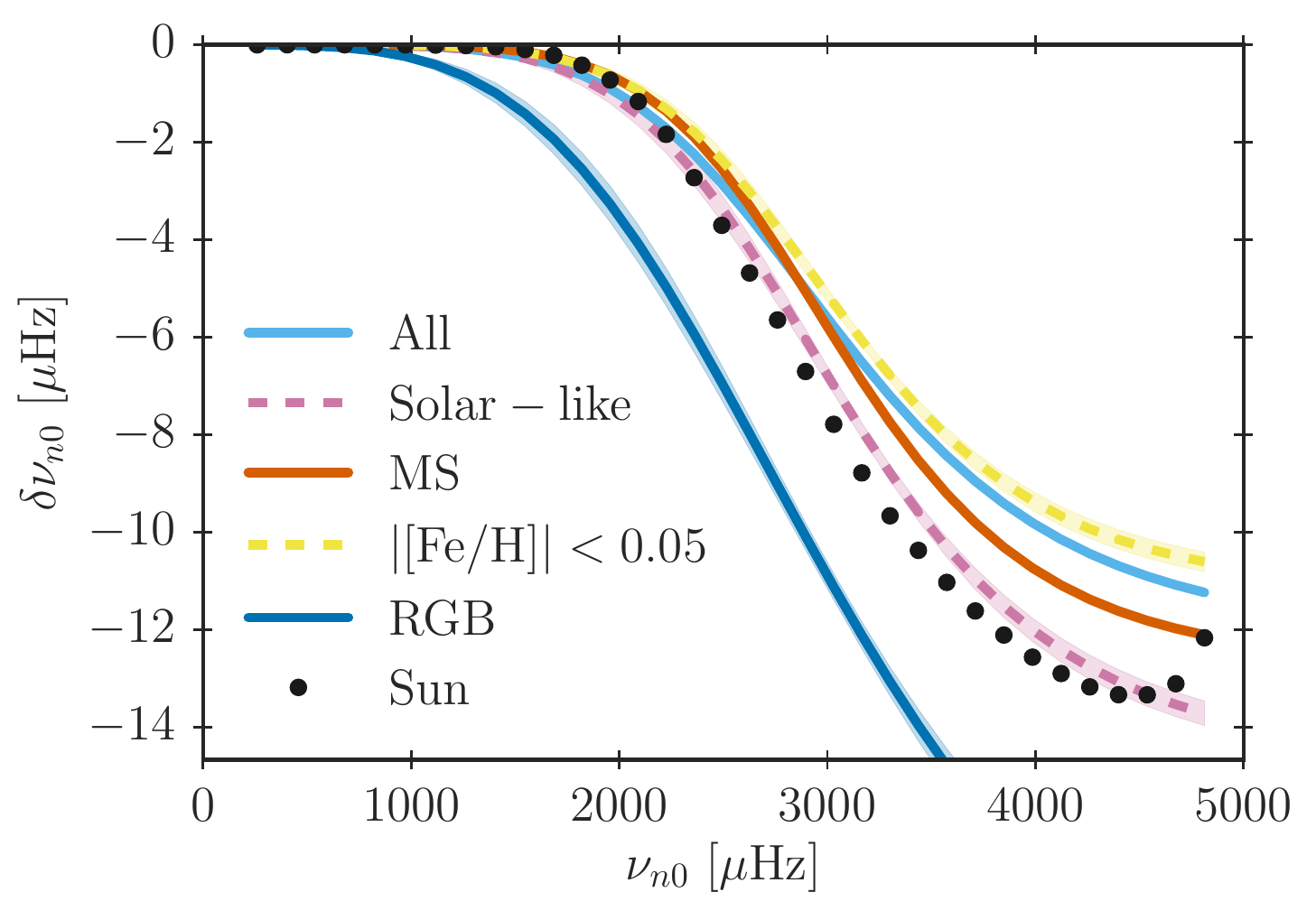}
\caption{Same as Fig.~\ref{fig:poly} but for the Lorentzian surface correction relation.
}
\label{fig:lorentz}
\end{figure}

{\color{black}As shown by \cite{Sonoi2015}, a Lorentzian fit that is derived directly from the solar model yields smaller residuals than the global fit presented in Fig.~\ref{fig:poly_lorentz}. Based on this notion combined with Fig.~\ref{fig:errorplane}, one may suspect a fit to a sample of solar-like dwarfs to perform better, when inferring the solar surface effect.
Indeed}, when recomputing the fit based on different subsamples of the grid and subsequently examining the solar case, a more detailed picture emerges: the performance of the fit depends on the subsample. This is illustrated in Fig.~\ref{fig:poly} and Fig.~\ref{fig:lorentz}. 

Besides including all 315 models in the fitting procedure, we have looked at the following subsamples:
\begin{enumerate}
\item 45 models for which $5500\,\mathrm{K}<T_\mathrm{eff}<5800\,\mathrm{K}$ and $\log g > 4.0$. There are no restrictions on $\mathrm{[Fe/H]}$. We refer to this sample as \textit{solar-like} stars.
\item 163 models for which $\log g > 4.0$. There are no restrictions $T_\mathrm{eff}$ on $\mathrm{[Fe/H]}$. We refer to this sample as \textit{MS} stars.
\item 56 models for which $-0.05<\mathrm{[Fe/H]}<0.05$. There are no restrictions $T_\mathrm{eff}$ on $\log g$.
\item 152 models for which $\log g < 4.0$. There are no restrictions $T_\mathrm{eff}$ on $\mathrm{[Fe/H]}$. We refer to this sample as \textit{RGB} stars.
\end{enumerate}

Figs~\ref{fig:poly} and ~\ref{fig:lorentz} include the uncertainties on the predicted frequency difference inferred from the fitting procedure, showing that the different fits are incompatible --- in some cases, the line width is broader than the inferred uncertainties. According to Figs~\ref{fig:poly} and ~\ref{fig:lorentz}, the best fit is obtained, when the models in the sample close match the target: the residuals become significantly lower, when we restrict ourselves to the 45 "solar-like" stars. The fact that different subsamples lead to different reconstructions of the solar case implies that a single scaling relation that applies to all evolutionary stages is not realizable with the presented approach. Also, it implies that previous studies, such as the inference of a surface correction relation by \cite{Sonoi2015}, is subject to a selection bias: in the mentioned case, the authors have selected ten PMs predominantly with $T_\mathrm{eff}>6000\,$K.

{\color{black}
For the computation of the fitting coefficients we have used interpolated $\langle \mathrm{3D} \rangle$-envelopes, since additional PMs add information to the analysis due the unique interior structure of the associated UPMs. Had we used only the original Stagger-grid simulations, we would have been limited by the small sample size. The lower number of PMs would impose restrictions on the subsample selection and affect the robustness of the inferred fitting coefficients. Furthermore, the distribution of simulations within the Stagger grid would introduce a selection bias. Interpolation gives us control over this selection bias.}


\section{Best-Fitting models} \label{sec:best-fitting}

{\color{black}
In this section, we take a closer look at the stellar parameters of two \textit{Kepler} stars, by building on the analysis presented above. As shown in Section~\ref{sec:Surfcorr}, the Lorentzian parametrization reliably encodes the structural surface effect, if the fit is based on models, whose global parameters resemble those of the target. Rather than computing further PMs, we hence use this parametrization. 

This approach has the advantage that other authors may easily reproduce and implement our results. Furthermore, using this approach, we can investigate the limitations of such parameterizations. For instance, in the previous section we fond that the performance of the Lorentzian fit depends on the subsample, upon which the fit is determined. Using our parametrization allows us to discuss what ramifications this has for the stellar parameter estimates, based on a comparison with \cite{Sonoi2015}. This is of interest, since surface correction relations are commonly used.

\subsection{Maximum Likelihood Estimation}}

Having established a parametrization of the structural surface effect and the associated constraints on the applicability hereof, we can use this parametrization to evaluate stellar parameters based on a grid search, by comparing the corrected model frequencies with observations. We do this by using maximum likelihood estimation (MLE). For this purpose, we construct a denser grid of un-patched stellar models and attribute a likelihood,
\begin{equation}
\mathcal{L} = \mathcal{L}_\mathrm{seis} \mathcal{L}_\mathrm{spec}, \label{eq:likelihood},
\end{equation}
to each model in the grid. The details of the grid are summarized in Subsection~\ref{sec:grid2}. 
In Eq.~(\ref{eq:likelihood}), $\mathcal{L}_\mathrm{seis}$ denotes the seismic contribution to the likelihood. For simplicity, we assume that the model frequencies are uncorrelated and that the uncertainties are Gaussian. While we note that both assumption are only approximately valid, we also note that at least the latter is commonly used \citep[e.g.][]{Aguirre2015, Nsamba2018}. The seismic contribution to the likelihood his hence
\begin{equation}
\mathcal{L}_\mathrm{seis} = \prod_i^N \frac{1}{\sqrt{2 \pi \sigma_i^2}} \exp \left( - \frac{(\nu_{\mathrm{corr},i}-\nu_{\mathrm{obs},i})^2}{2\sigma_i^2} \right). \label{eq:seislike}
\end{equation}
Here $\nu_{\mathrm{corr},i}$ is the $i$th corrected model frequency of the un-patched model, for which we determine the likelihood. In the following, we use Eq.~(\ref{eq:lorentz}), i.e. our Lorentzian correction, employing fitting parameters that were determined based on an appropriate subregion of the Kiel diagram. {\color{black} The subregion was determined using the grid presented in Section~\ref{sec:Surfcorr} --- that is, the models from Section~\ref{sec:Surfcorr} that enter the fit span the same region of the parameter space as the denser grid presented in Subsection~\ref{sec:grid2} does. We elaborated upon this in the next Subsection.
} 

In order to be consistent with the derivation of the frequency correction relation, we only include radial frequencies ($\ell = 0$) in our analysis. For comparison, we additionally repeat the analysis using the surface correction relation by \cite{Kjeldsen2008}, \cite{Ball2014} and \cite{Sonoi2015}. 

In Eq.~(\ref{eq:seislike}), each of the $N$ observed frequencies ($\nu_{\mathrm{obs},i}$) is compared to the corresponding corrected model frequency {\color{black}--- the number $N$ is determined by the number of observed radial modes and will hence be fixed for a given star.}
We allow for the radial order of the model frequencies to deviate from observations by up to 2, since the radial orders of the observed frequencies may be incorrectly assigned. We choose the attribution of the radial order that leads to the lowest frequency deviation for the lowest frequency. Finally, in Eq.~(\ref{eq:seislike}), $\sigma_i$ denotes the uncertainty of the frequency difference. We note that this is not simply the uncertainty of the observed frequency, since the corrected frequencies are themselves a function of the observed frequencies. In order to take this into account, we use Eq.~(\ref{eq:jakobian}). As we assume the observations to be uncorrelated and the associated uncertainties to be Gaussian, Eq.~(\ref{eq:jakobian}) reduces to the absolute value of the derivative of $\nu_{\mathrm{obs},i}-\nu_{\mathrm{corr},i}$ with respect to $\nu_{\mathrm{obs},i}$ times the uncertainty on $\nu_{\mathrm{obs},i}$. We consistently ignore the uncertainties on the fitting parameters here, since these are not available for all surface corrections. 

When computing the likelihood, we also take spectroscopic constraints on the effective temperature and the metallicity into account. Again, we assume the uncertainties to be Gaussian:
\begin{equation}
\mathcal{L}_\mathrm{spec} = \frac{1}{2 \pi \sigma_T \sigma_\mathrm{[Fe/H]}} \exp \left(-\frac{\Delta T_\mathrm{eff}^2}{2\sigma_T^2} -\frac{\Delta \mathrm{[Fe/H]}^2}{2\sigma_\mathrm{[Fe/H]}^2} \right).
\end{equation}
Here $\sigma_T$ and $\sigma_\mathrm{[Fe/H]}$ denote the standard deviation of the spectroscopically deduced effective temperature and metallicity, respectively. $\Delta T_\mathrm{eff}$ and $\Delta \mathrm{[Fe/H]}$ denote the difference between the observationally constrained value and the model prediction for each of the two quantities.

Out of convenience, we work with the natural logarithm of the likelihood, when performing calculations. {\color{black}More specifically,} we work with the averaged log-likelihood:
\begin{equation}
\langle \ln \mathcal{L} \rangle= \frac{1}{N+2}\ln \mathcal{L} \label{eq:loglike},
\end{equation}
{\color{black}for our $N$ frequencies and two spectroscopic observables.}
The stellar model that {\color{black}maximizes} the quantity in Eq.~(\ref{eq:loglike}) is considered the best-fitting model in the grid.

In order to ascribe a measure for the goodness of fit, we provide $\chi^2$ of the best-fitting model:
\begin{equation}
\chi^2 = \frac{1}{N+2} \left( \sum_i^N \frac{(\nu_{\mathrm{corr},i}-\nu_{\mathrm{obs},i})^2}{\sigma_i^2} + \frac{\Delta T_\mathrm{eff}^2}{\sigma_T^2} + \frac{\Delta \mathrm{[Fe/H]}^2}{\sigma_\mathrm{[Fe/H]}^2} \right).
\end{equation}

The parameter values of the best-fitting model do not on their own encode comprehensive information about the likelihood function. We hence likewise provide the weighted mean and the weighted variance of each parameter: having $K$ stellar models in the grid, the weighted mean of a parameters $\beta$ is:
\begin{equation}
\langle \beta \rangle = \frac{\sum_k^K \beta_k \langle \mathcal{L}_k\rangle}{\sum_k^K \langle \mathcal{L}_k\rangle}.
\end{equation}
Here $\langle\mathcal{L}_k\rangle$ is the exponential of the quantity in Eq.~(\ref{eq:loglike}) for the $k$the model. Equivalently, the weighted variance is 
\begin{equation}
\left( \sigma_\beta^\mathrm{mod} \right)^2 = \frac{\sum_k^K (\beta_k-\langle \beta \rangle)^2 \langle \mathcal{L}_k\rangle}{\sum_k^K \langle \mathcal{L}_k\rangle}.
\end{equation}
A weighted mean that deviates from the parameter value of the best-fitting model and a large variance imply a likelihood function that deviates from a single narrow peak.


\subsection{The stellar model grid} \label{sec:grid2}

While a few hundred models are sufficient for the purpose of robustly establishing the parametrizations in Section~\ref{sec:Surfcorr}, a higher sampling of the parameter space is needed, when evaluating stellar parameters based on a grid-search. Using the same input physics as in Subsection~\ref{sec:grid1}, we therefore computed a new grid with a higher sampling rate: we computed 4,007 stellar evolution tracks with masses between $0.8\,\mathrm{M}_\odot$ and $1.3\,\mathrm{M}_\odot$ in steps of $0.01\,\mathrm{M}_\odot$. This includes tracks with the same composition as the underlying solar calibration as well as initial compositions on the pre-MS that correspond to a metallicity between $\mathrm{[Fe/H]}=-0.20$ and 0.55 in steps of 0.01. Due to microscopic diffusion the metallicity decrease with age and can hence be lower than the listed initial values. {\color{black}We note that the high mass stars will be more efficient at draining metals from their photosphere, which implies that the sampling in metallicity depends on the other global parameters: as discussed in Section~\ref{sec:grid1}, the depletion maybe artificially high at higher masses. For our current analysis, however, this is of no concern, since the Sun and the investigated stars all have masses below $1.2\,\mathrm{M}_\odot$.  } 

From this set of evolution tracks we selected a dense grid of stellar structure models. Since we have already dealt with the solar case, we investigate a different region of the Kiel diagram. Applying the same selection criteria as listed in Subsection~\ref{sec:grid1} but demanding at least two model frequencies to be evaluated above $500\,\mu$Hz, we selected 150,719 un-patched models with effective temperatures between $5255\,$K and $6120\,$K, $\log g$ between 3.78 and 4.30 and metallicities between -0.42 and 0.47. Due to the constraints set by theoretical evolution tracks the models are irregularly distributed in the parameter space spanned by $T_\mathrm{eff}$,$\log g$, and $\mathrm{[Fe/H]}$. Radial frequencies below $500\,\mu$Hz are ignored.

We have derived the coefficients for our Lorentzian surface correction relation based on an equivalent region of the Kiel diagram: the surface correction is hence determined using 73 models from the grid presented in Subsection~\ref{sec:grid1}, for which $T_\mathrm{eff}$, $\log g$ and $\mathrm{[Fe/H]}$ lie in the intervals $[5257\,\mathrm{K}, 6050\,\mathrm{K}]$, $[4.05,4.28]$ and $[-0.49,0.38]$, respectively. The coefficients are summarized in Table~\ref{tab:fitval2}.
\begin{table*}
	\centering
	\caption{Pendant to Table~\ref{tab:fitval}, containing fitting parameters based on a global fit to 73 PMs in our target region.}
	\label{tab:fitval2}
	\begin{tabular}{lccccccccccccccccccccc} 
		\hline
		  & A & B & C & D\\		
		\hline  
$\alpha$ &$(-1.644\pm 0.029)\times 10^{-3}$ & $8.507\pm 0.181$ & $-1.825\pm 0.025$ &  $(8.154\pm 0.1.132)\times 10^{-1}$   \\
$\beta$ &$(1.839\pm  0.129)\time10^{1}$ & $-7.056\pm 0.665$ & $1.616\pm 0.094$ & $(-7.966\pm 0.432)\times 10^{-1}$  \\		\hline
	\end{tabular}
\end{table*}


\subsection{Comparisons}

We evaluate the best-fitting models in our grid using the MLE procedure based on a Lorentzian fit to PMs in the relevant region of the Kiel diagram as described above. For comparison we repeat the evaluation of the best-fitting stellar parameters, employing surface correction relations by other authors. We briefly elaborate on these below.

\cite{Kjeldsen2008} suggest Eq.~(\ref{eq:poly}), i.e. a fit to a power law, as the functional form of the surface correction, setting $b=4.90$ and 
\begin{equation}
a = \frac{\langle \nu_{\mathrm{obs},i}\rangle - s \langle \nu_i\rangle}{N^{-1} \sum_{i=1}^N \left[\nu_{\mathrm{obs},i}/\nu_\mathrm{max} \right]}.
\end{equation}
Here $s=1$, $\nu_{\mathrm{obs},i}$ denotes the $i$th observed frequency and $\nu_{i}$ denotes the corresponding model frequency. In the following, we will refer to the surface correction relation by \cite{Kjeldsen2008} as K08.

\cite{Sonoi2015} suggest to use a surface correction with functional form given by Eq.~\ref{eq:lorentz}, i.e. a Lorentzian fit. The parameters $\alpha$ and $\beta$ are based on ten PMs and do not include a dependence on metallicity:
\begin{equation}
\log |\alpha| = - 7.69 \log T_\mathrm{eff} - 0.629 \log g -28.5,
\end{equation}
and $\alpha$ is negative, while
\begin{equation}
\log \beta = - 3.86 \log T_\mathrm{eff} + 0.235 \log g + 14.2.
\end{equation}
We will refer to this surface correction as S15. A comparison between the results obtained, when using our own surface correction relation, and those obtained, when using S15, in principle allows for a discussion on the importance of the inclusion of metallicity. However, different results may mainly reflect the selection bias.

Rather than computing coefficients for an assumed functional form using PMs or solar observations, \cite{Ball2014} present a surface correction relation that has been deduced from the variational principle by \cite{Gough1990} based on physical considerations:
\begin{equation}
\frac{\delta \nu_i}{\nu_\mathrm{ac}} = \mathcal{I}^{-1} \left( a_{-1}\frac{ \nu_{\mathrm{ac}}}{\nu} + a_3 \frac{\nu^3}{\nu_{\mathrm{ac}}^3} \right) \label{eq:ball}.
\end{equation}
Here $\mathcal{I}$ is the mode inertia \citep[cf.][]{jcd2008}, while $\nu_\mathrm{ac}$ denotes the acoustic cutoff frequency:
\begin{equation}
\frac{\nu_\mathrm{ac}}{\nu_{\mathrm{ac}\odot}}=\frac{\nu_\mathrm{max}}{\nu_{\mathrm{max}\odot}}, \label{eq:vac}
\end{equation}
where we set $\nu_{\mathrm{ac}\odot}=5000\,\mu\mathrm{Hz}$ in accordance with \cite{Ball2014}. We will refer to this surface correction relation as B\&G.

The coefficients $a_{-1}$ and $a_3$ are given by 
\begin{equation}
\begin{pmatrix}
a_{-1} \\
a_3
\end{pmatrix} = (\mathbfss{X}^\mathrm{T}\mathbfss{X})^{-1} \mathbfss{X}^\mathrm{T} \mathbf{y}, \label{eq:a1a3}
\end{equation}
where $\mathbf{y}$ denotes the frequency difference between the observed frequencies and the model frequencies in units of the uncertainty of the observed frequencies ($\sigma_i^\mathrm{obs}$)
\begin{equation}
y_i = \frac{\nu_{\mathrm{obs},i} - \nu_i}{\sigma_i^\mathrm{obs}}
\end{equation}
and the matrix $\mathbfss{X}$ is a combination of model frequencies and the associated mode inertias 
\begin{equation}
X_{i,1} = \frac{\nu_\mathrm{ac}^2 \nu_i^{-1}}{\mathcal{I}_i\sigma_i^\mathrm{obs}}, \quad X_{i,2} = \frac{\nu_\mathrm{ac}^{-2} \nu_i^{3}}{\mathcal{I}_i\sigma_i^\mathrm{obs}}. \label{eq:X12}
\end{equation}
Our surface correction relation was inferred from PMs without taking modal effects into account. Assuming that the asymptotic analysis that underlies B\&G includes all relevant contributions, Eq.~(\ref{eq:ball}) is not subject to this inadequacy. Hence, a comparison between the results obtained when using our surface correction relation, and those obtained when using Eq.~(\ref{eq:ball}), can be used to determine how the neglect of modal effects alters the determination of model parameters. To a certain extent, this also holds true for the comparison with K08.

Since both K08 and B\&G rely on the observed frequencies in order to determine the surface correction, these two surface correction relations have limited predictive power in comparison to our parameterization of the structural contribution of the surface effect or to the one by \cite{Sonoi2015}. The fact that both \cite{Kjeldsen2008} and \cite{Ball2014} adjust the fitting coefficients according to data also implies that a comparison with our surface correction or S15 does not take place on equal footings. In order to investigate the ramifications hereof, we have repeated the evaluation of the best-fitting model based on B\&G fixing the coefficients to the values obtained for our solar calibration model: $a_{-1} = 4.27\times 10^{-12}$ and $a_3 = -6.93\times 10^{-11}$. In this evaluation of the coefficients, we have used observed BiSON frequencies \citep{Broomhall2009,Davies2014}. We refer to the use of the solar values for the coefficients as ``case b'', while we refer to the use of Eq.~(\ref{eq:a1a3}) as ``case a''. 

We note that the incorrect modeling of the surface layers results in wrong boundary conditions for solving the stellar structure equations throughout the predicted evolution. This leads to systematic errors in the associated stellar parameters, which none of the surface correction relations take into account. We refer to \cite{Mosumgaard2018} for a discussion on how the implementation of information from 3D simulations into 1D models alters the evolution tracks.

Furthermore, we recall that we restrict ourselves to radial modes ($\ell = 0$) in all cases, in order to facilitate a meaningful comparison between all surface corrections.


\subsection{Hare and Hound}

In order to evaluate the performance of our Lorentzian surface correction relation across the parameter space, we test how well the relation is able to recover the stellar parameters of PMs. For comparison we have repeated the same hare and hound exercise, using K08, S15 and B\&G, cases a and b. 

Since the frequencies and global parameters of the PMs have no intrinsic uncertainties as such, we have attributed artificial uncertainties to mimic observational constraints. Hence, we assume the uncertainty on the effective temperature and metallicity to be $100\,$K and $0.1$, respectively. All frequencies have been attributed an uncertainty of $0.3\,\mu$Hz. {\color{black}For comparison, the symmetric errors on the radial modes vary between $0.07\,\mu$Hz and $1.36\,\mu$Hz, for the two \textit{Kepler} stars presented in Subsection~\ref{sec:kepler}. While the chosen value is hence certainly not unreasonable, we note that the use of constant uncertainties is a simplifying approximation: for instance, the uncertainty will tend to reflect the mode amplitude.}

We have performed five hare and hound exercises. {\color{black}
We have chosen models with $T_\mathrm{eff}$ between 5600$\,$K and 5900$\,$K and mostly with low metallicities. The reason for this selection is the attempt to reliably sample the likelihood for all surface corrections: especially the use of K08 leads to metallicities that are systematically too high and effective temperatures that are much lower than the target value, as discussed below.
The global parameters of the five target models for the hare and hound exercises are furthermore selected so that we explore the region of the parameter space that is expected to be relevant for the \textit{Kepler} stars that are to be addressed in Subsection~\ref{sec:kepler}. 
}
The results obtained for the hare and hound exercises from the MLE are summarized for each surface correction in Tables~\ref{tab:HareHound} and \ref{tab:HareHound2}.

\begin{table*}
	\centering
	\caption{Global parameters from the hare and hound exercise based on PMs. Further parameters are listed in Table~\ref{tab:HareHound2}. The first column contains an identifier. The second through the fourth column contain the global parameters of the PMs. The fifth column lists the used surface correction relation. The sixth through the eighth column summaries the parameter values for the best-fitting model, the weighted mean and the square root of the weighted variance. The last column contains the initial metallicity of the best-fitting model. {\color{black}The purpose of including the latter quantity is solely to demonstrate that none of the models are at the edge of our grid of stellar models; hence we do not provide uncertainties on the initial metallicity.} We compare the surface corrections by Ball \& Gizon (2014) in combination with Eq.~(\ref{eq:a1a3}) (B\&G, a), the one by Ball \& Gizon (2014) with fixed coefficients (B\&G, b), the one presented in this paper, the one by Kjeldsen et al (2008) (K08), and the one by Sonoi et al. (2015) (S15).}
	\label{tab:HareHound}
	\begin{tabular}{lccccccccccccccccccccc} 
		\hline
		 Hare & $T_\mathrm{eff} $ & $\log g$ & $\mathrm{[Fe/H]}$ & Surf. Corr. & $T_\mathrm{eff}^\mathrm{mod}$ & $\log g^\mathrm{mod}$ & $\mathrm{[Fe/H]}^\mathrm{mod}$ & $\mathrm{[Fe/H]}_\mathrm{init}^\mathrm{mod}$   \\
         & K & dex & dex & & K & dex & dex & dex \\
		\hline
        A & $5624$ & $4.054$ & $-0.028$& K08 & $5372$, $5378\pm31$ & $4.045$, $4.046\pm0.002$ & $0.081$, $0.105\pm0.046$ & $0.20$ \\    
        & & & & B\&G (a) & $5596$, $5613\pm183$ & $4.053$, $4.053\pm0.008$ & $0.020$, $-0.037\pm0.119$ & $0.13$   \\
        & & & & B\&G (b) & $5463$, $5487\pm59$ & $4.045$, $4.045\pm0.005$ & $0.008$, $-0.024\pm0.072$ & $0.13$  \\
          & & & & S15 & $5596$, $5543\pm101$ & $4.053$, $4.049\pm0.009$ & $0.002$, $-0.028\pm0.071$ & $0.13$ \\
& & & & This work & $5553$, $5604\pm109$ & $4.045$, $4.049\pm0.009$ & $-0.086$, $-0.076\pm0.093$ & $0.04$ \\
          \hline
B & $5926$ & $4.185$ & $-0.301$ & K08 & $5486$, $5467\pm91$ & $4.181$, $4.182\pm0.003$ & $0.096$, $0.141\pm0.068$ & $0.25$ \\
& & & & B\&G (a) & $6077$, $6016\pm90$ & $4.197$, $4.196\pm0.004$ & $-0.334$, $-0.274\pm0.092$ & $-0.14$ \\
        & & & & B\&G (b) & $5917$, $5944\pm47$ & $4.182$, $4.188\pm0.006$ & $-0.344$, $-0.285\pm0.072$ & $-0.16$ \\
          & & & & S15 & $5963$, $5944\pm58$ & $4.194$, $4.196\pm0.004$ & $-0.238$, $-0.183\pm0.069$ & $-0.06$ \\
& & & & This work & $5934$, $5944\pm44$ & $4.189$, $4.191\pm0.004$ & $-0.258$, $-0.249\pm0.069$ & $-0.08$ \\
          \hline
C & $5895$ & $4.053$ & $-0.262$& K08 & $5399$, $5426\pm51$ & $4.035$, $4.036\pm0.004$ & $-0.049$, $-0.057\pm0.055$ & $0.07$\\
& & & & B\&G (a) & $5878$, $5792\pm171$ & $4.056$, $4.052\pm0.008$ & $-0.197$, $-0.175\pm0.123$ & $-0.04$ \\
        & & & & B\&G (b) & $5508$, $5617\pm121$ & $4.027$, $4.035\pm0.009$ & $-0.266$, $-0.247\pm0.054$ & $-0.14$ \\
          & & & & S15 & $5953$, $5867\pm79$ & $4.053$, $4.053\pm0.006$ & $-0.349$, $-0.249\pm0.099$ & $-0.17$ \\
& & & & This work & $5819$, $5900\pm74$ & $4.042$, $4.052\pm0.007$ & $-0.343$, $-0.289\pm0.059$
& $-0.19$ \\
          \hline        
D &$5781$ & $4.055$ & $0.06$ & K08 & $5412$, $5398\pm53$ & $4.041$, $4.037\pm0.006$ & $0.213$, $0.159\pm0.114$ & $0.33$ \\
& & & & B\&G (a) &$5668$, $5681\pm181$ & $4.049$, $4.049\pm0.008$ & $0.090$, $0.060\pm0.115$ & $0.22$\\
        & & & & B\&G (b) & $5667$, $5578\pm114$ & $4.036$, $4.033\pm0.009$ & $-0.099$, $-0.072\pm0.092$ & $0.03$ \\
          & & & & S15 & $5710$, $5740\pm63$ & $4.051$, $4.051\pm0.005$ & $0.076$, $0.037\pm0.083$ & $0.21$\\
& & & & This work & $5813$, $5758\pm84$ & $4.051$, $4.053\pm0.006$ & $-0.042$, $0.042\pm0.118$ & $0.06$\\
          \hline        
E & $5874$ & $4.219$ & $-0.008$& K08 & $5486$, $5460\pm68$ & $4.213$, $4.210\pm0.004$ & $0.322$, $0.313\pm0.046$ & $0.47$ \\
& & & & B\&G (a) & $5995$, $5844\pm183$ & $4.225$, $4.222\pm0.004$ & $-0.075$, $0.061\pm0.166$ & $0.09$ \\
        & & & & B\&G (b) & $5874$, $5924\pm64$ & $4.219$, $4.217\pm0.002$ & $-0.008$, $-0.089\pm0.083$ & $0.15$\\
          & & & & S15 & $5831$, $5834\pm36$ & $4.224$, $4.224\pm0.001$ & $0.099$, $0.100\pm0.049$ & $0.25$\\
& & & & This work & $5886$, $5897\pm23$ & $4.220$, $4.219\pm0.001$ & $-0.019$, $-0.032\pm0.025$ & $0.14$ \\
          \hline
	\end{tabular}
\end{table*}

\begin{table*}
	\centering
	\caption{Pendant to Table~\ref{tab:HareHound} with other stellar parameters. The last column contains the $\chi^2$ of the best-fitting model. The identifiers of the PMs correspond to those in Table~\ref{tab:HareHound}.}
	\label{tab:HareHound2}
	\begin{tabular}{lccccccccccccccccccccc} 
		\hline
		 Hare & $M$ & $R$ & $t$ & Surf. Corr. & $M^\mathrm{mod}$ & $R^\mathrm{mod}$ & $t^\mathrm{mod}$ & $\chi^2$\\
         & $\mathrm{M}_\odot$ & $\mathrm{R}_\odot$ & Gyr & & $\mathrm{M}_\odot$ & $\mathrm{R}_\odot$ & Gyr & \\
		\hline
         A & $0.999$ & $1.557$ & $10.8$ & K08 & $0.959$, $0.97\pm0.015$ & $1.541$, $1.546\pm0.008$ & $13.57$, $13.26\pm0.48$ & $16.62$\\
         & & & & B\&G (a) & $0.999$, $0.992\pm0.045$ & $1.557$, $1.552\pm0.021$ & $10.88$, $11.09\pm2.14$ & $0.4$ \\
        & & & & B\&G (b) & $1.000$, $0.954\pm0.029$ & $1.54$, $1.537\pm0.015$ & $12.82$, $12.79\pm1.01$  & $12.64$  \\
          & & & & S15 & $1.000$, $0.972\pm0.051$ & $1.557$, $1.543\pm0.026$ & $10.88$, $12.06\pm2.05$ & $2.49$  \\
& & & & This work & $0.949$, $0.972\pm0.052$ & $1.532$, $1.543\pm0.025$ & $12.35$, $11.56\pm1.89$ & $1.64$ \\
          \hline
B & $0.950$ & $1.305$ & $9.77$ & K08 & $0.950$, $0.960\pm0.019$ & $1.311$, $1.315\pm0.008$ & $13.64$, $13.5\pm1.34$ & $11.59$ \\
& & & & B\&G (a) & $1.000$, $0.998\pm0.018$ & $1.32$, $1.321\pm0.007$ & $7.50$, $7.98\pm0.85$ & $6.29$ \\
        & & & & B\&G (b) & $1.000$, $0.964\pm0.033$ & $1.295$, $1.310\pm0.014$ & $10.35$, $9.31\pm1.00$ & $13.32$ \\
          & & & & S15 & $1.000$, $1.005\pm0.021$ & $1.319$, $1.326\pm0.009$ & $8.53$, $8.34\pm0.65$ & $14.21$\\
& & & & This work & $0.970$, $0.978\pm0.022$ & $1.312$, $1.315\pm0.01$ & $9.22$, $8.95\pm0.61$ & $8.53$ \\
          \hline
C &$0.999$ & $1.559$ & $8.86$& K08 & $0.919$, $0.925\pm0.022$ & $1.526$, $1.529\pm0.012$ & $14.37$, $14.0\pm0.89$ & $10.84$ \\
& & & & B\&G (a) & $1.019$, $1.001\pm0.041$ & $1.568$, $1.56\pm0.019$ & $8.61$, $9.61\pm1.80$ & $0.36$ \\
        & & & & B\&G (b) & $1.000$, $0.911\pm0.052$ & $1.497$, $1.518\pm0.026$ & $14.85$, $12.93\pm2.17$ & $12.05$ \\
          & & & & S15 & $1.000$, $0.996\pm0.036$ & $1.55$, $1.556\pm0.018$ & $8.67$, $9.15\pm0.94$ & $6.38$\\
& & & & This work & $0.939$, $0.993\pm0.038$ & $1.530$, $1.555\pm0.018$ & $10.55$, $9.00\pm1.19$ & $3.41$ \\
          \hline
D & $1.099$ & $1.63$ & $7.78$& K08 & $1.029$, $1.004\pm0.041$ & $1.604$, $1.591\pm0.021$ & $11.36$, $12.07\pm1.17$& $16.47$\\
& & & & B\&G (a) & $1.069$, $1.064\pm0.041$ & $1.618$, $1.614\pm0.018$ & $8.92$, $9.07\pm1.72$& $0.97$ \\
        & & & & B\&G (b) & $1.000$, $0.973\pm0.057$ & $1.581$, $1.573\pm0.029$ & $10.46$, $11.58\pm2.04$& $18.04$ \\
          & & & & S15 & $1.000$, $1.074\pm0.033$ & $1.623$, $1.619\pm0.016$ & $8.50$, $8.46\pm0.78$& $5.19$ \\
& & & & This work & $1.069$, $1.083\pm0.039$ & $1.615$, $1.622\pm0.019$ & $8.19$, $8.22\pm0.90$
 & $5.09$ \\
          \hline
E & $1.05$ & $1.319$ & $7.67$& K08 & $1.030$, $1.015\pm0.02$ & $1.316$, $1.311\pm0.008$ & $10.69$, $11.41\pm1.18$ & $6.28$ \\        
        & & & & B\&G (a) & $1.070$, $1.067\pm0.018$ & $1.323$, $1.324\pm0.007$ & $6.51$, $7.49\pm1.28$ & $2.18$ \\
        & & & & B\&G (b) & $1.000$, $1.034\pm0.016$ & $1.319$, $1.312\pm0.007$ & $7.67$, $7.69\pm0.42$ & $19.53$ \\
          & & & & S15 & $1.000$, $1.081\pm0.011$ & $1.330$, $1.331\pm0.005$ & $7.25$, $7.20\pm0.23$ & $6.23$\\
& & & & This work & $1.050$, $1.048\pm0.009$ & $1.318$, $1.317\pm0.003$ & $7.59$, $7.57\pm0.25$ & $2.76$ \\
          \hline
\end{tabular}
\end{table*}

We find our parametrization of the surface effect to recover the stellar parameters rather well, and hence we infer that the Lorentzian fit encodes detailed information about the structural contribution to the surface effect.

S15 has been calibrated on fewer models and with different input physics than our PMs, and a dependence of the coefficients on metallicity was not included. Nevertheless, it likewise performs rather well, when establishing the model parameters in the hare and hound exercise. This may reflect the fact that S15 resembles our surface correction relation in  being a parametrization of the structural contribution only. Also, this underlines that a Lorentzian parametrization captures the structural contribution.

In contrast, K08 becomes less reliable the further we move from the solar case. When employing this surface correction, the effective temperature of the PM is systematically underestimated by a few hundred degrees. {\color{black}Also, the inferred metallicity is much higher than for the original model.}

\begin{figure}
\centering
\includegraphics[width=\linewidth]{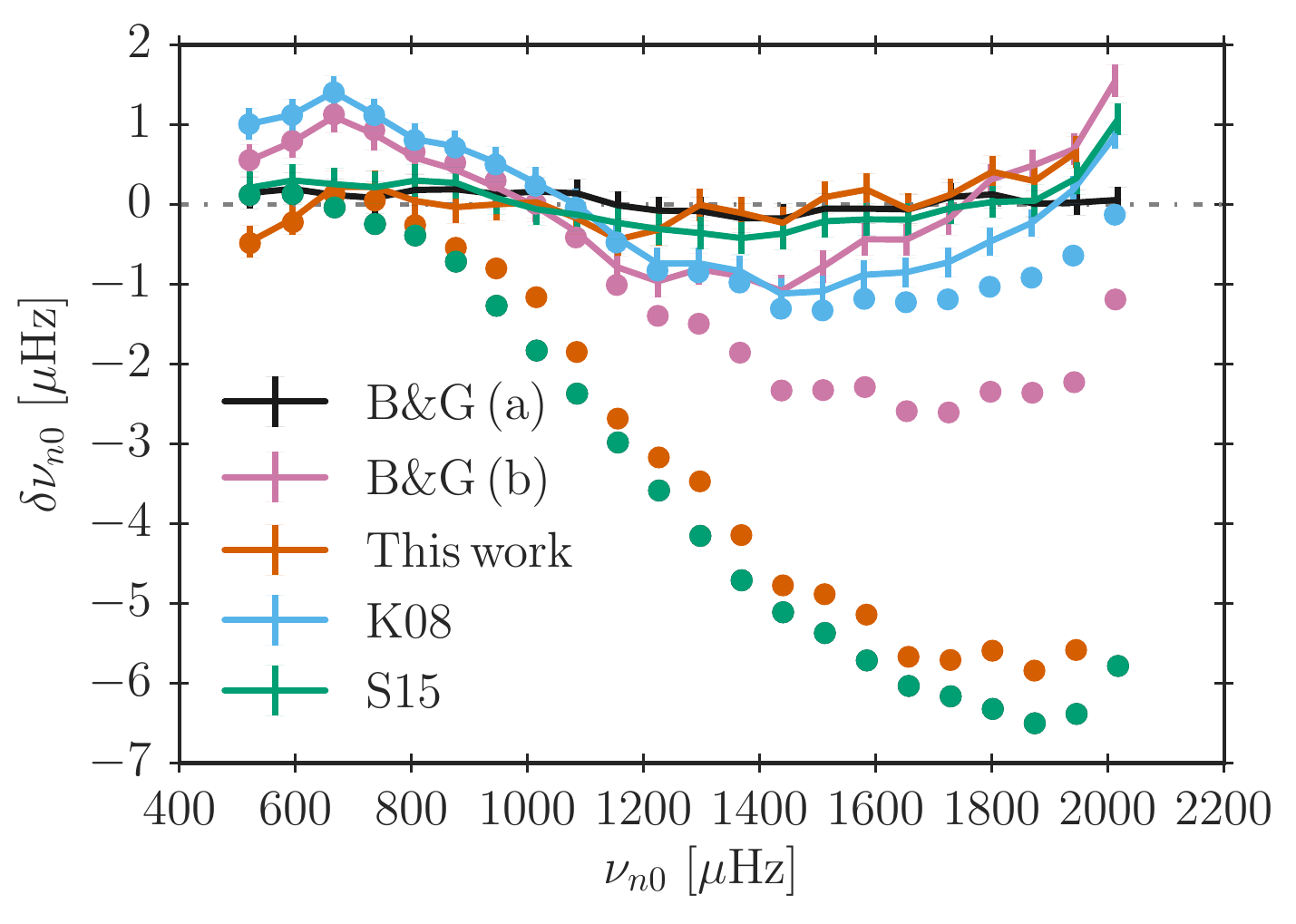}
\caption{Frequency difference between model frequencies and observations for the PM with the identifier A in Tables~\ref{tab:HareHound} and \ref{tab:HareHound2}, using different surface correction methods. The solid lines show the corrected frequencies. The circular markers in the same color show the associated un-corrected frequencies of the respective best-fitting models.
}
\label{fig:M095Fm00}
\end{figure}

Using K08 leads to the selection of a model, whose frequencies already closely match the frequencies of the PM without the necessity of a surface correction. This peculiarity also occasionally occurs, when using B\&G with fixed coefficient (case b), but contrasts the selection obtained with the other surface correction relations, including B\&G in combination with Eq.~(\ref{eq:a1a3}) (case~a). It is exemplified in Fig.~\ref{fig:M095Fm00}.

As discussed in the next subsection, K08 performs better when confronted with \textit{Kepler} data. Therefore, the selection by K08 may partly reflect the neglect of modal effects in our surface corrections relation, i.e. the fact that the frequency difference between the UPM and the associated PM does not fit the expected functional form: as shown in Section~\ref{sec:Surfcorr}, the difference between a PM and the associated UPM is not well-described by a power law. 

Furthermore, in the case of B\&G, the inferred surface correction is lower when the coefficient are fixed to the solar values (case~b) than when Eq.~(\ref{eq:a1a3}) is used to establish the coefficients (case~a) for each model anew. It hence stands to reason that the evaluation of the coefficients based on the solar case distorts the model selection in the investigated region of the parameter space. This conclusion seems to agree with the qualitative finding of Subsection~\ref{sec:SolRec}: no unique set of coefficient values are applicable throughout the Kiel diagram. This agrees with the conclusions drawn by \cite{Ball2017}: surface corrections that employ coefficients that are calibrated to the Sun perform comparatively poorly, when fitting \textit{Kepler} observations.

Despite the neglect of the modal effects in the computation of the model frequencies of the PMs, B\&G performs rather well, when using Eq.~(\ref{eq:a1a3}) (case~a). {\color{black}However, in case~a, B\&G also leads to the broadest probability distributions, as $a_{-1}$ and $a_3$ are treated as free parameters. Thus, the higher accuracy comes at the price of lower precision.}

While the different surface corrections recover parameters such as the effective temperature and the metallicity with varying success, the gravitational acceleration is always accurately and precisely reproduced. This stems from the fact that we mainly constrain our models through the oscillation frequencies: the global oscillation properties of stars are determined by both the mass ($M$) and radius ($R$) of the star. Thus, both the frequency of maximum amplitude and the large frequency separation are proportional to $g$ and $\sqrt{g}$, respectively.


\subsection{\textit{Kepler} stars} \label{sec:kepler}

Having established how well our surface corrections reproduce the stellar parameters of PMs in a grid search, we proceed to an analysis of observational data. We search for best-fitting models using our MLE approach for the \textit{Kepler} stars KIC~10516096 (Manon) and KIC~5950854. The observational constraints on the employed radial frequencies as well as on $T_\mathrm{eff}$ and $\mathrm{[Fe/H]}$ are taken from \cite{Lund2017}. Both stars are in the LEGACY dwarf sample. The obtained stellar parameters from our MLE approach are summarized in Tables~\ref{tab:KIC105} and \ref{tab:KIC595}, respectively. 

{\color{black}An analysis of main-sequence stars with global properties that are rather similar to those of the Sun have already been presented in \JJ. We therefore selected KIC~10516096 and KIC~5950854, in order to investigate stars at a later evolutionary stage. Furthermore, spectroscopic constraints indicate that the global parameters of these two stars lie well within our computed grid of stellar models, which allows us to properly map the probability distributions.}
\begin{table*}
	\centering
	\caption{Best-fitting parameters for KIC~10516096. According to spectroscopy, $T_\mathrm{eff} = 5964\pm77$ and $\mathrm{[Fe/H] = -0.11\pm0.10}$, respectively.}
	\label{tab:KIC105}
	\begin{tabular}{lccccccccccccccccccccc} 
		\hline
        Quantity & Unit & K08 & B\&G (a) & B\&G (b) & S15 & This work \\
        \hline
        $\chi^2$ & & $5.73$ & $2.45$ & $7.06$ & $6.44$ & $7.93$ \\
       	$T_\mathrm{eff}$ & K & $5535$, $5592\pm63$ & $5887$, $5912\pm151$ & $5772$, $5724\pm68$ & $5612$, $5657\pm81$ & $5760$, $5734\pm65$ \\
        $\log g$ & dex & $4.169$, $4.17\pm0.004$ & $4.178$, $4.179\pm0.005$ & $4.175$, $4.173\pm0.006$ & $4.173$, $4.173\pm0.006$ & $4.175$, $4.173\pm0.005$ \\
        $\mathrm{[Fe/H]}$ & dex  & $0.376$, $0.312\pm0.075$ & $0.104$, $0.074\pm0.143$ & $0.191$, $0.211\pm0.073$ & $0.336$, $0.274\pm0.084$ & $0.181$, $0.186\pm0.099$ \\
        $\mathrm{[Fe/H]}_\mathrm{init}$ & dex & 0.52 & 0.26 & 0.34 & 0.48 & 0.33 \\
        $M$ & $\mathrm{M}_\odot$ &  $1.080$, $1.081\pm0.027$ & $1.120$, $1.118\pm0.026$ & $1.110$, $1.101\pm0.04$ & $1.100$, $1.097\pm0.04$ & $1.100$, $1.090\pm0.029$ \\
        $R$ & $\mathrm{R}_\odot$ & $1.417$, $1.417\pm0.011$ & $1.428$, $1.425\pm0.01$ & $1.427$, $1.424\pm0.016$ & $1.423$, $1.421\pm0.015$ & $1.421$, $1.417\pm0.012$ \\
        $t$ & Gyr & $9.272$, $8.979\pm0.957$ & $6.592$, $6.507\pm1.038$ & $7.369$, $7.852\pm1.248$ & $8.347$, $8.244\pm1.247$ & $7.641$, $7.995\pm0.751$ \\
        \hline
	\end{tabular}
\end{table*}

\begin{table*}
	\centering
	\caption{Best-fitting parameters for KIC~5950854. According to spectroscopy, $T_\mathrm{eff} = 5853\pm77$ and $\mathrm{[Fe/H] = -0.23\pm0.10}$, respectively.}
	\label{tab:KIC595}
	\begin{tabular}{lccccccccccccccccccccc} 
		\hline
        Quantity & Unit & K08 & B\&G (a) & B\&G (b) & S15 & This work \\
        \hline
        $\chi^2$ & & $2.08$ & $0.6$ & $2.33$ & $3.22$ & $3.4$ \\
       	$T_\mathrm{eff}$ & K & $5702$, $5694\pm170$ & $5861$, $5757\pm175$ & $5673$, $5647\pm148$ & $5650$, $5584\pm129$ & $5656$, $5671\pm145$ \\
        $\log g$ & dex & $4.23$, $4.235\pm0.013$ & $4.234$, $4.218\pm0.034$ & $4.234$, $4.234\pm0.012$ & $4.235$, $4.232\pm0.010$ & $4.231$, $4.234\pm0.011$ \\
        $\mathrm{[Fe/H]}$ & dex & $-0.117$, $-0.037\pm0.161$ & $-0.211$, $-0.134\pm0.158$ & $-0.020$, $0.007\pm0.159$ & $0.000$, $0.019\pm0.153$ &  $-0.054$, $-0.033\pm0.151$ \\
        $\mathrm{[Fe/H]}_\mathrm{init}$ & dex & 0.05 & -0.04 & 0.14 & 0.16 & 0.11 \\
        $M$ & $\mathrm{M}_\odot$ & $0.930$, $0.963\pm0.077$ & $0.950$, $0.949\pm0.063$ & $0.960$, $0.963\pm0.072$ & $0.960$, $0.944\pm0.060$ & $0.940$, $0.955\pm0.065$ \\
        $R$ & $\mathrm{R}_\odot$ & $1.226$, $1.238\pm0.031$ & $1.232$, $1.256\pm0.054$ & $1.24$, $1.24\pm0.029$ & $1.238$, $1.232\pm0.024$ & $1.231$, $1.236\pm0.026$ \\
        $t$ & Gyr & $11.985$, $11.356\pm3.495$ & $9.971$, $11.219\pm3.066$ & $11.281$, $11.702\pm3.337$ & $11.472$, $12.79\pm2.799$ & $12.078$, $11.691\pm3.061$ \\
        \hline
	\end{tabular}
\end{table*}

As can be seen from the tables, the use of B\&G in combination with Eq.~(\ref{eq:a1a3}) (case~a) leads to parameter estimates that are in good agreement with the spectroscopic constraints. Lesser agreement is achieved, when using the other surface correction relations and the same relation in case b. Thus, with the exception of B\&G case~a, the evaluated effective temperature is systematically too low, while the metallicity is too high in most cases.

The stellar age estimates vary strongly between the different fits, while estimates of $\log g$ and the stellar radius are mutually consistent and precisely determined. The mass estimates also closely match. Again, this seems to reflect the fact that the global oscillation properties encode stringent constraints on $g$. 

\begin{figure}
\centering
\includegraphics[width=\linewidth]{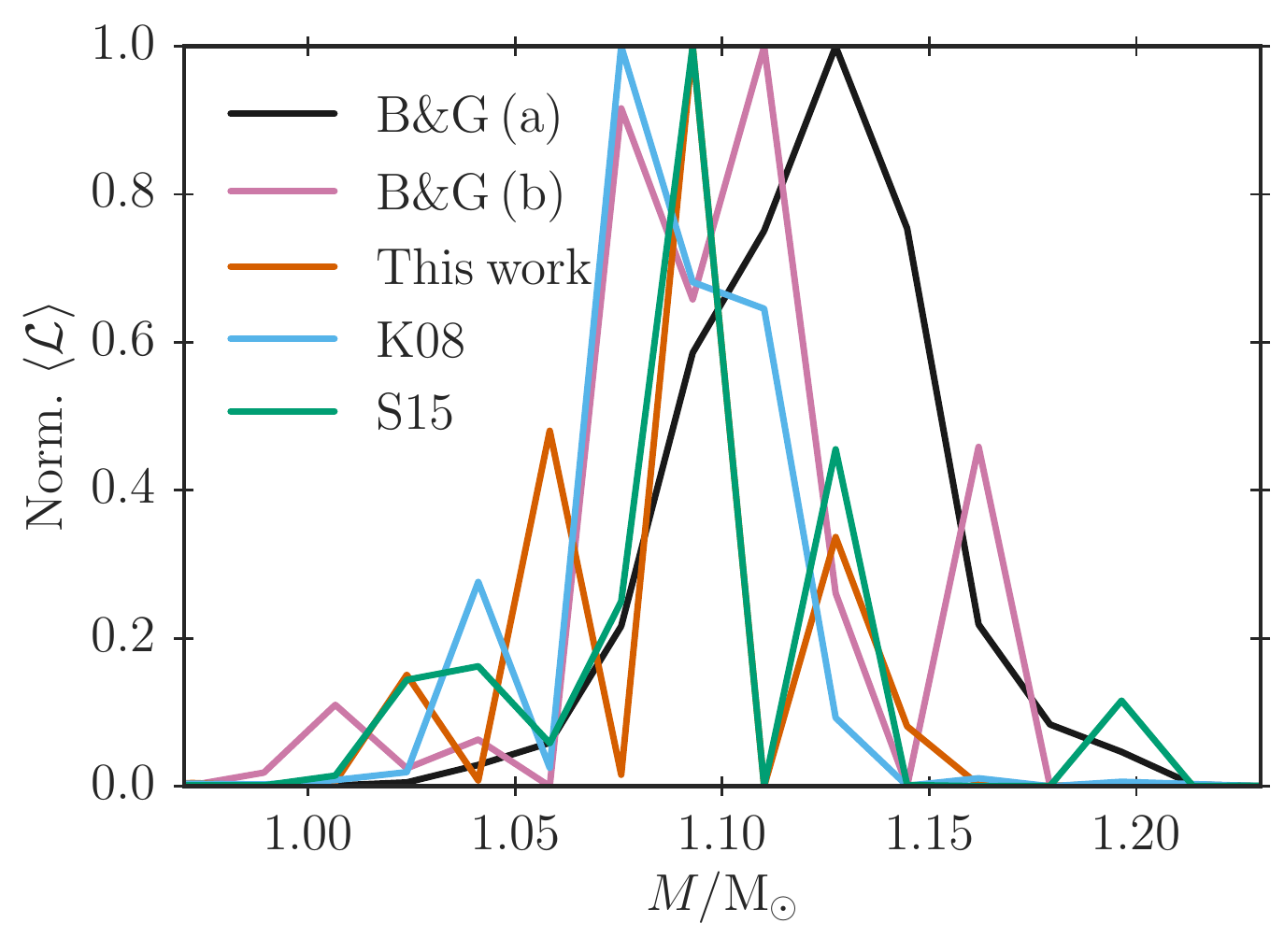}
\caption{Projection of the normalized exponential of the averaged log-likelihood for the stellar mass for KIC~10516096. The normalization factor is established in such a way that the likelihood of the maximum likelihood model is 1. We use 30 bins, smoothing the likelihood function.
}
\label{fig:Mass105}
\end{figure}

We include the projected likelihood for the mass of KIC~10516096 in Fig.~\ref{fig:Mass105}. As can be seen from the figure, the use of the surface correction relation by B\&G, case~a, leads to a much broader distribution than the remaining surface correction relations do. This is a general feature that we find for all projections --- for both \textit{Kepler} stars as well as in the hare and hound exercise. This can be explained by the fact that Eq.~(\ref{eq:a1a3}) adjusts the coefficients in order to optimize the fit, {\color{black}i.e. the fit has two additional free parameters}, which ascribes artificially high likelihoods to incorrect structures. 

\begin{figure}
\centering
\includegraphics[width=\linewidth]{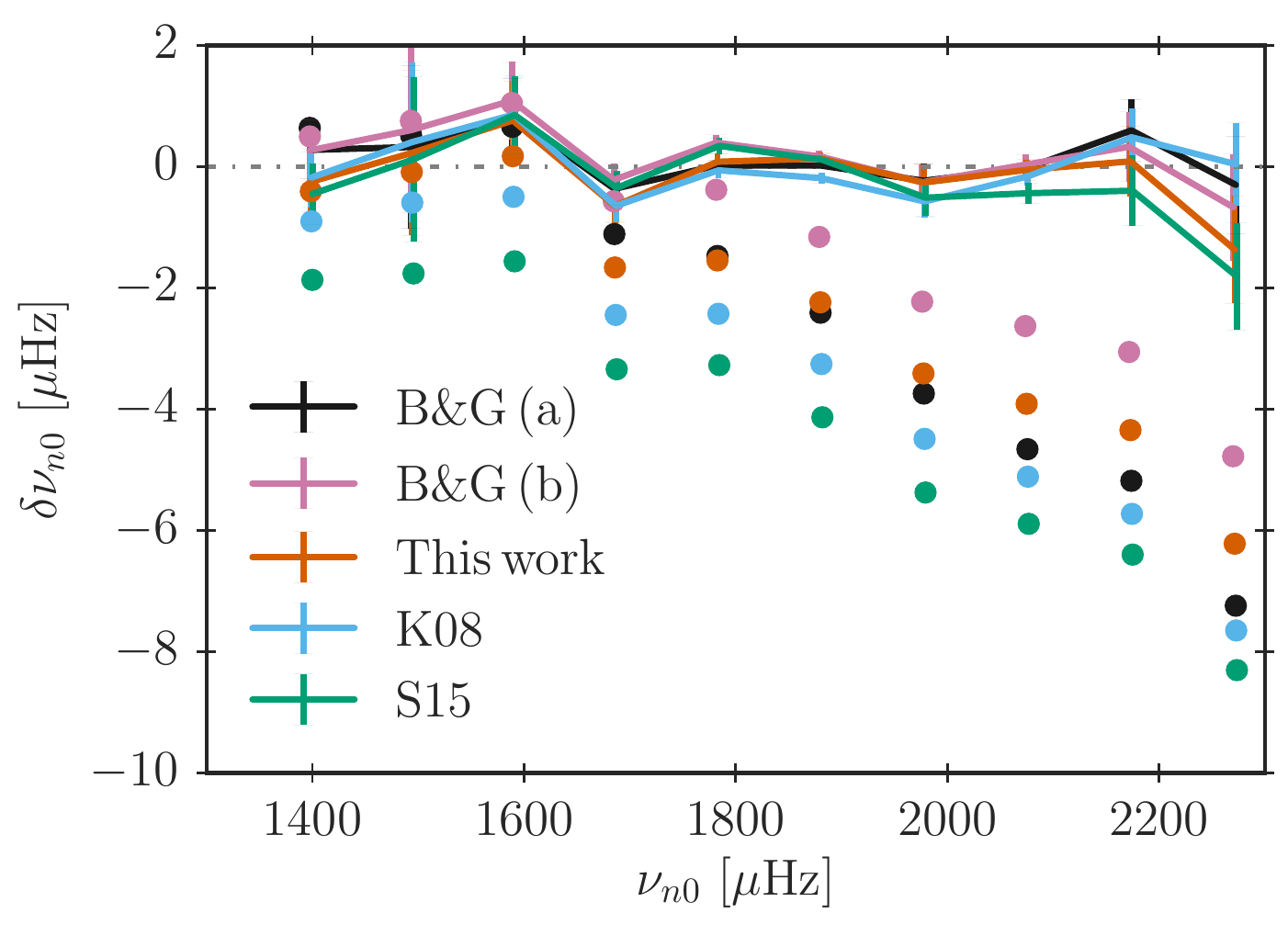}
\caption{Same as Fig.~\ref{fig:M095Fm00} but for KIC~5950854.
}
\label{fig:Frek595}
\end{figure}

\begin{figure}
\centering
\includegraphics[width=\linewidth]{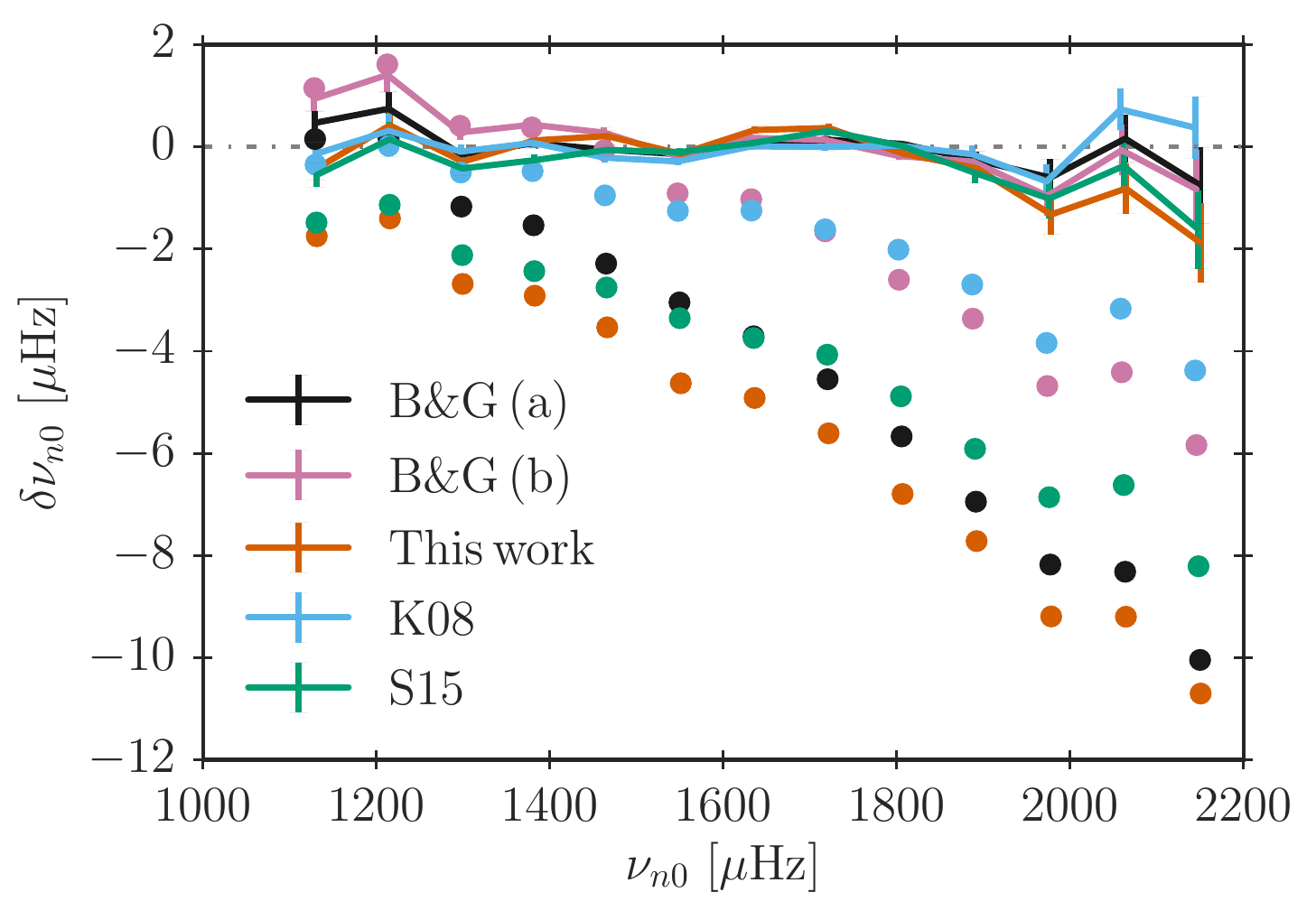}
\caption{Same as Fig.~\ref{fig:M095Fm00} but for KIC~10516096.
}
\label{fig:Frek105}
\end{figure}

The frequency difference between the obtained best-fitting models and observations are shown in Figs~\ref{fig:Frek595} and \ref{fig:Frek105}. We note that the surface correction relations that rely on solar coefficients generally predict a smaller frequency correction than the remaining surface correction relations do.

At this point, we refer to \cite{Nsamba2018} for another recently published determination of the systematics arising from different surface correction methods. \cite{Nsamba2018} determine posterior probability distributions for the stellar mass, radius and age using different surface correction relations by employing a Markov Chain Monte Carlo algorithm. They then compare the evaluated parameter values with those obtained, when following the same Bayesian inference technique but using frequency ratios suggested by \cite{Roxburgh2003}. Since these frequency ratios have been shown to be rather insensitive to the surface layers, the associated parameter values are a suitable reference value and are taken as a measure of the systematic errors introduced by the use of each surface correction relation. \cite{Nsamba2018} conclude that B\&G reproduces the mass, radius and age that is inferred from Frequency ratios. The same conclusion was drawn by \cite{Basu2018}. {\color{black}This suggests that B\&G correctly encodes the structural as well as the modal effects}. Thus, we may use parameter values that are based on case a of B\&G as a benchmark for the other surface correction relations.

According to \cite{Nsamba2018}, K08 leads to larger intrinsic systematics for the mass, radius and age than B\&G does. This is compatible with our results: for both \textit{Kepler} stars K08 leads to lower mass estimates and a higher stellar age. Furthermore, \cite{Nsamba2018} conclude that S15 performs rather well. Again, this agrees with our mass, radius and age estimates: S15 is rather consistent with case a of B\&G. This being said, as regards S15, a direct comparison between our results and those obtained by \cite{Nsamba2018} is misleading, since \cite{Nsamba2018} defines the coefficients differently: $\beta$ is set to 4.0, while $\alpha$ calibrated based on the observed frequencies.

\subsection{Patched models of \textit{Kepler stars}}

\begin{figure}
\centering
\includegraphics[width=\linewidth]{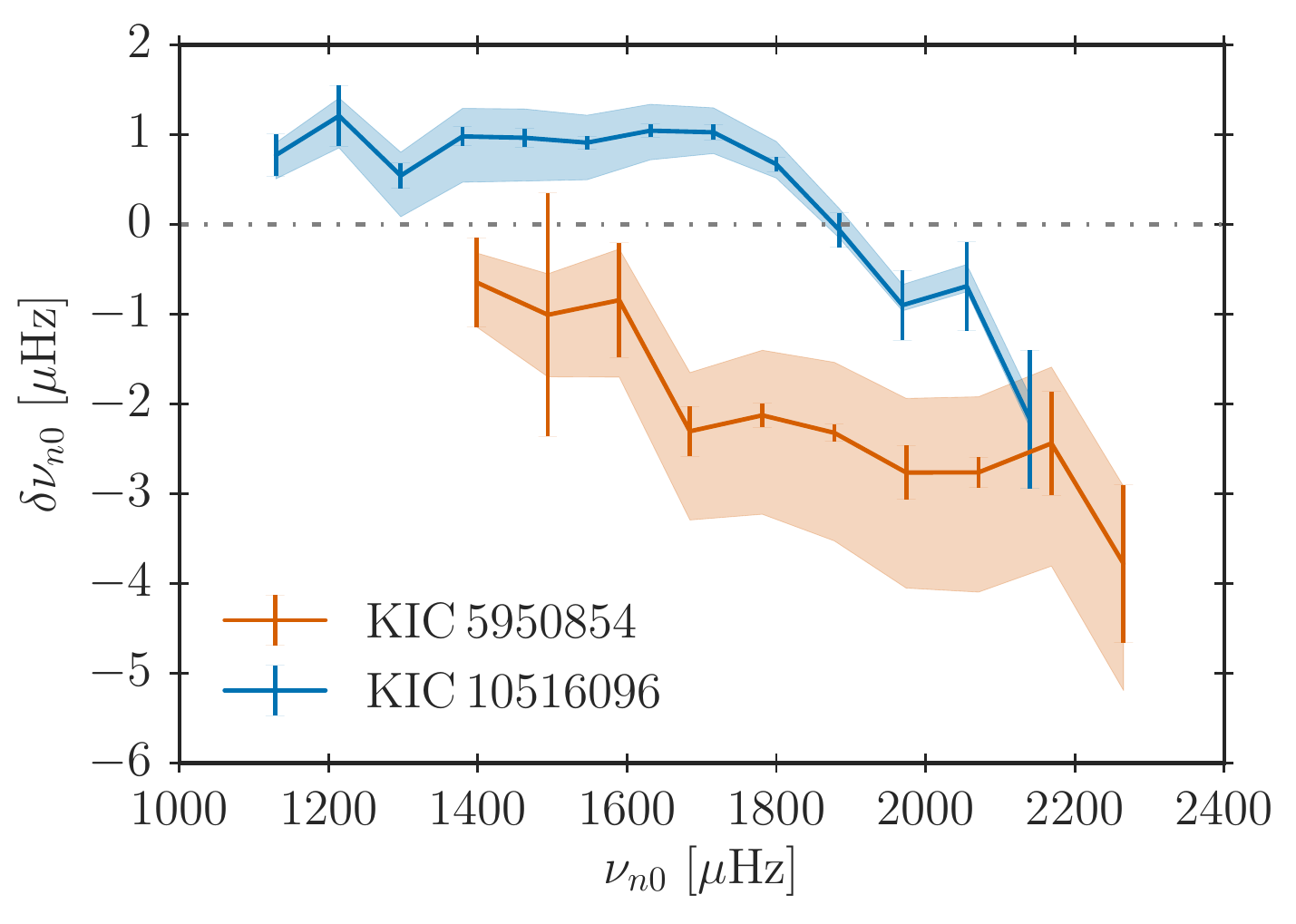}
\caption{Frequency difference between observations and PM for KIC~5950854 and KIC~10516096. The UPM corresponds to the best-fitting model established using case a of B\&G. The solid lines show the frequencies that are obtained, when using $P$ as the patching quantity. The shaded area shows the frequency shift that is introduced, when using $\rho$ or $T$ as the patching quantity. The error bars illustrates the error on the observed frequencies. 
}
\label{fig:Frek_BG}
\end{figure}

Having established that case a of B\&G is plausibly the most reliable surface correction relation for establishing stellar parameters, we can use the best-fitting UPMs obtained based on B\&G to construct PMs, in order to model the structure of the \textit{Kepler} targets. The associated difference between the adiabatic model frequencies of the PMs and the observations are summarized in Fig.~\ref{fig:Frek_BG}.

\begin{figure}
\centering
\includegraphics[width=\linewidth]{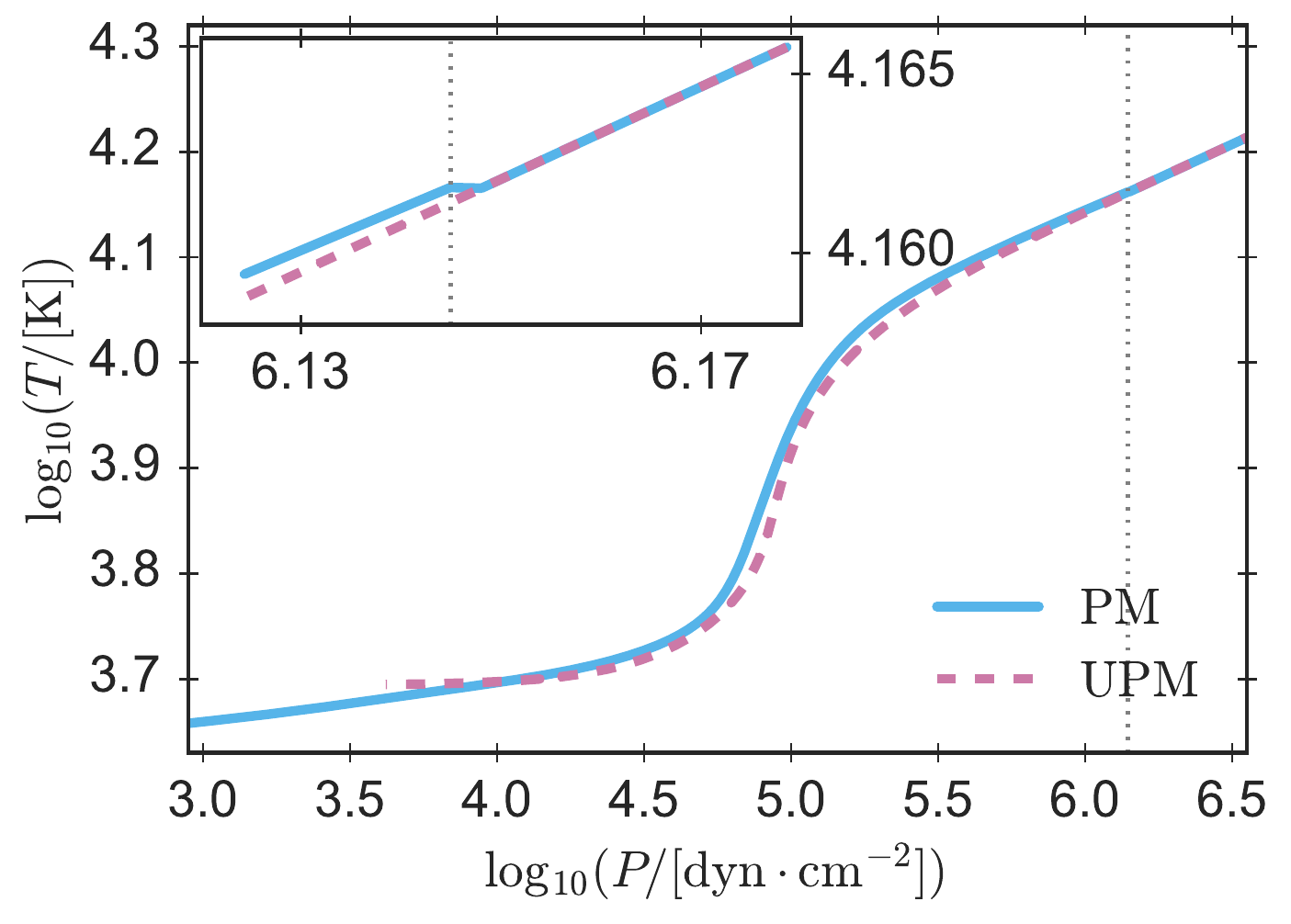}
\caption{\color{black}Temperature as a function of pressure for KIC~10516096. The cyan solid line and the dashed magenta line show the stratification of the UPM and PM, respectively. The dotted gray line shows the location of the innermost mesh point of the patched envelope. A zoom-in is included in the upper left corner. For the PM, $\rho$ has been used as the patching quantity. Consequently, there is discontinuity in the temperature stratification at the patching point.
}
\label{fig:PT105}
\end{figure}

{\color{black}In order to demonstrate the systematic errors that are introduced by our choice of the patching quantity, Fig.~\ref{fig:Frek_BG} includes three PMs for each \textit{Kepler} star, using $P$, $\rho$ or $T$ as the respective patching quantities. As elaborated upon in Section~\ref{sec:Surfcorr}, the sensitivity of the oscillation frequencies to the patching quantity stem from discontinuities in the patched structure. This is illustrated in Fig.~\ref{fig:PT105} for KIC~10516096. In this figure, we use the density as the patching quantity, which results in a discontinuity in the temperature and pressure stratifications: while $\rho(r)$ is a smooth function by construction, neither $T(r)$ nor $P(r)$ is --- consequently, $T(P)$ in Fig.~\ref{fig:PT105} is neither. One way to quantify this discontinuity is to evaluate the change in the stellar radius that is introduced by a change in the patching quantity: using the temperature rather than the density as the patching quantity would shift the radius of the patching point by $18\,$km. To put this into perspective, we note that the radius of both stars is larger than that of the Sun (cf. Tables~\ref{tab:KIC105} and \ref{tab:KIC595}) and that patch is performed roughly $2\times 10^3\,$km below the surface in either case. In comparison, this shift would be $73\,$km in the case of KIC~5950854. As a result, the oscillation frequencies of KIC~5950854 are more sensitive to the choice of the patching quantity than those of KIC~10516096, as shown in Fig.~\ref{fig:Frek_BG}.
}

According to \cite{Houdek2017}, the modal surface effect counteracts the structural surface effect, increasing the model frequencies --- that is at least the case for the Sun. At first glance, Fig.~\ref{fig:Frek_BG} seems to contradict this conclusion in the case of KIC~5950854: the obtained model frequencies of the PM are systematically higher than the observed frequencies. However, in accordance with \cite{Ball2016}, \cite{Sonoi2015} and \JJ, we use $\Gamma_1$, when calculating model frequencies, while \cite{Houdek2017} modify $\Gamma_1$ by the gas pressure as fraction of the total pressure. This is done in order to consistently take the contribution of the turbulent pressure into account and alters the model frequencies \citep[cf.][]{Joergensen2018}. Thus, the frequency difference between the model frequencies and observations in Fig.~\ref{fig:Frek_BG} do not only include contributions that \cite{Houdek2017} denote as modal. 

If we assume that case a of B\&G accurately reproduces the stellar structure and frequencies, Fig.~\ref{fig:Frek_BG} rather points towards the conclusion that the structural surface effect alone does not reliably mimic the total surface effect: the deviations are quite significant for KIC~5950854. {\color{black}This is an interesting notion, since it illustrates, why the surface correction in this paper as well as the correction relation by \cite{Sonoi2015} is subject to systematic errors, when estimating stellar parameters.}

Whether B\&G indeed accurately reproduces the underlying stellar structures is not settled. To answer this question, a detailed analyses of how the structural as well as the modal effect depends on the stellar parameters is needed. This is beyond the scope of this paper. This being said, the comparison by \cite{Nsamba2018} and \cite{Basu2018}, in which they infer the systematic error introduced by the B\&G performs relatively well.

We note that the model frequencies in Fig.~\ref{fig:Frek_BG} may also reflect inconsistencies in the physical assumptions that enter the UPMs and the patched $\langle \mathrm{3D} \rangle$-envelopes, which would affect the model frequencies. Moreover, due to the restrictions imposed by the interpolation scheme, the patched envelopes may be too shallow, i.e. the patch may not have been performed sufficiently deep within the adiabatic region for the oscillation frequencies to be entirely insensitive to the patching depth \citep[cf.][]{Joergensen2017}. In order to exclude this scenario, deeper 3D simulations are required, which again lies beyond the scope of the present paper.

Finally, interpolation errors may give rise to further frequency shifts: the structure of the patched envelope may be incorrect. As pointed out in \JJ, the sampling of existing grids of 3D envelopes may hence be too low in some regions of the parameters space. To investigate this possibility, an extension of the Stagger grid is needed. 


\section{Conclusion} \label{sec:concl}

We present a method that allows for the interpolation of mean stratifications of 3D simulations of stellar envelopes in metallicity. Hereby, we add an additional parameter to the interpolation scheme published in \JJ, broadening its applicability. The interpolated envelopes can be used to correct the improper structure of current 1D stellar structure models by substituting these layers. We present such patched models for both the Sun and stars in the \textit{Kepler} field.

We show that our interpolation scheme reproduces the correct mean stratification quite accurately throughout the parameter space. Based on the solar case, we note that interpolation errors as well as the limited depth of the interpolated structure may lead to a frequency shift of a few microhertz. In order to improve the performance of the interpolation scheme, a refinement of the Stagger grid of 3D envelopes is needed. Also, our results call for deeper 3D simulations.

Based on our interpolation scheme, we derive a parameterizations of the structural contribution of the surface effect, using the same functional forms as suggested by \cite{Sonoi2015}: a power law and a Lorentzian. We find the coefficients of the parameterizations to be sensitive to the selected sample of stellar models. Hence we conclude that no unique set of coefficients is applicable for all evolutionary stages. Consequently, our results discourages the use of any semi-empirical surface correction that is calibrated to fit the Sun for stars, whose parameters deviate strongly from the solar case. Rather, surface corrections must be calibrated based on stars that closely resemble the target. 

Our parameterizations only encode the structural surface effect, neglecting modal contributions \cite[cf.][]{Houdek2017}. Using a simple maximum likelihood estimation approach, we derive stellar parameters using both our parametrization and the surface correction by \cite{Ball2014} that incorporates structural as well as modal effects. From this comparison, we conclude that the neglect of modal effects alter the parameter estimates. Consequently, without further adjustments, our results do not support the use of our parameterization or the surface correction by \cite{Sonoi2015}, when evaluating stellar parameters.

As shown by \cite{Joergensen2018}, it is possible to avoid the necessity of post-evolutionary patching, by including mean 3D envelopes at each time step, adjusting the boundary conditions of the stellar structure equations accordingly. Having shown that we are able to reliably interpolate in metallicity, we hence plan to implement an interpolation in the stellar composition into our stellar evolution code. In this way, we can take the metallicity dependence of structure into account on the fly. Including this additional information from 3D simulations, may possibly alter the evolution tracks \citep[cf.][]{Mosumgaard2018}. 

\section*{Acknowledgements}

We thank J. Christensen-Dalsgaard for his kind assistance with \textsc{adipls} and record our gratitude to Z. Magic and R. Trampedach for supplying the employed mean 3D hydrodynamic simulations as well as for fruitful discussions and helpful insights. Furthermore, we thank R. Collet for sharing his invaluable understanding of the Stagger grid and for his generous assistance. We likewise acknowledge T. En{\ss}lin, C.~L. Sahlholdt and J.~R. Mosumgaard for the useful discussions and input. V.S.A. acknowledges support from VILLUM FONDEN (research grant 10118) and the Independent Research Fund Denmark (Research grant 7027-00096B). {\color{black}Finally, we record our gratitude to our referee, who provided very detailed feedback that showed an exceptional level of care and effort and that improved the paper in many ways.}











\bsp	
\label{lastpage}
\end{document}